\documentclass[traditabstract]{aa}
\usepackage{txfonts}
\usepackage{graphicx}
\usepackage{natbib}

\begin{document}


\title{NGC\,6240: merger-induced star formation \& gas dynamics\thanks{Based on 
    observations at the European Southern Observatory VLT (079.B-0576).}}

\author{H. Engel \inst{1} \and R. I. Davies \inst{1} \and R. Genzel\inst{1} \and L. J. Tacconi\inst{1} \and E. K. S. Hicks\inst{1} \and E. Sturm\inst{1} \and T. Naab\inst{2} \and P. H. Johansson\inst{2} \and S. J. Karl\inst{2} \and C. E. Max\inst{3} \and A. Medling\inst{3} \and P. P. van der Werf\inst{4}}

\institute{}
	\institute{Max Planck Institut f\"ur extraterrestrische Physik, Postfach 1312,
  85741 Garching, Germany
\and
	Universit\"atssternwarte, Scheinerstrasse 1, 81679 M\"unchen, Germany
\and
	Center for Adaptive Optics, University of California, 
  Santa Cruz, CA 95064, USA
\and 
Leiden Observatory, Leiden University, P.O. Box 9513, NL-2300 RA Leiden, NL}

\offprints{H. Engel \\ \email{hauke@mpe.mpg.de}}

   \date{}

\abstract{
We present spatially resolved integral field spectroscopic K-band data at a
resolution of 0.13\arcsec\ (60\,pc) and interferometric CO(2-1) line
observations of the prototypical merging system \object{NGC\,6240}.
Despite the clear rotational signature, the stellar kinematics in the
two nuclei are dominated by dispersion.
We use Jeans modelling to derive the masses and the mass-to-light ratios
of the nuclei. 
Combining the luminosities with the spatially resolved
Br$\gamma$ equivalent width shows that only 1/3 of
the K-band continuum from the nuclei is associated with the most
recent star forming episode; and that less than 30\% of the
system's bolometric luminosity and only 9\% of its stellar mass is due to this
starburst. The star formation properties, calculated from typical
merger star formation histories, demonstrate the impact of different
assumptions about the star formation history. 
The properties of the nuclei, and the existence of a prominent old
stellar population, indicate that the nuclei are remnants of the
progenitor galaxies' bulges.
}

\keywords{
galaxies: active ---
galaxies: individual (NGC\,6240) ---
galaxies: interactions --- 
galaxies: kinematics and dynamics --- 
galaxies: starburst ---
infrared: galaxies} 

   \authorrunning{H. Engel et al.}

\maketitle

\section{Introduction}
\label{sec:intro}
Major mergers are key drivers of galaxy evolution. Profoundly
transformational events, they substantially affect virtually all
properties of a galaxy. 
Mergers between disc galaxies are
believed to be responsible for triggering galaxy-wide
starbursts
\citep{toomre72,mih96,genzel98a,genzel98b,veilleux02,2005MNRAS.361..776S,hopkins06}),
quasar activity
\citep{sanders88a,sanders88b,sanders96,veilleux02,jogee04,2005MNRAS.361..776S,dasyra06b,hopkins06,hopkins08},
and the creation of elliptical galaxies
\citep{toomre72,toomre77,korm92,mih96,genzel01,dasyra06a,dasyra06b}.

Star formation activity in merging gas-rich galaxies peaks between
first peripassage and final coalescence
\citep{sanders96,mih96,veilleux02,2005MNRAS.361..776S}.
Due to absorption and re-emission by dust,
almost all of this energy is emitted at infrared wavelengths, giving rise to
extremely high infrared luminosities. Objects with $L_{\rm IR}$\,=\,10$^{11-11.9}L_\odot$ are called Luminous Infrared Galaxies (LIRGs), those with $L_{\rm IR}$\,$\geq$\,10$^{12}L_\odot$ UltraLuminous InfraRed Galaxies (ULIRGs). The local LIRG population consists of both (typically early-stage) mergers and non-interacting galaxies \citep{sanders92,alonso06}, whereas local ULIRGs are almost always interacting galaxies beyond first encounter \citep{sanders88a,sanders96,veilleux02,jogee04}. Local ULIRGs are predominantly powered by the starburst, except for the most luminous objects
\citep{genzel98b}.

Despite their recognised importance, we still lack a good
understanding of merger processes.
This is due both to the complexity and the shortlived nature of these
events.
A rare opportunity to study the transient phase between first
encounter and final coalescence of two merging gas-rich spirals is
afforded by NGC\,6240, at a distance of 97\,Mpc. 
With $L_{\rm IR}$\,$\sim$\,10$^{11.8}L_\odot$
\citep{sanders96}, it falls
just short of being formally classified as a ULIRG. \textit{HST} images show a
large-scale morphology dominated by tidal arms characteristic for
mergers past first peripassage \citep{gerssen04}. 
It is therefore more characteristic of the ULIRG class, to which it is
commonly assigned; and our analysis shows its luminosity is likely to
exceed the ULIRG threshold in the next 100--300\,Myr.
Two distinct nuclei, with a projected separation of
$\sim$1.5\arcsec\ or 700\,pc, are seen at infrared,
optical, and radio wavelengths \citep{tecza00,max05,max07,gerssen04,bes01}. Each
nucleus is host to an AGN, detected in hard X-rays \citep{kom03} and
at 5\,GHz \citep{gallimore04}. 
However, \cite{lut03} and \cite{arm06} estimate that the AGN
contribute less than half, and perhaps only 25\%, of the luminosity.
\cite{tac99} find the cold gas, as traced by
CO(2-1) emission, to be concentrated between the two nuclei. 
The 1-0\,S(1) H$_2$
line emission is one of the most powerful found in any galaxy to date
\citep{joseph84}, probably excited in shocks
\citep{werf93,sug97,tecza00,lut03}.
The stellar kinematics \citep{tecza00} display
rotation centred on each of the two nuclei, and extraordinarily large
stellar velocity dispersions ($\sim$350\,km\,s$^{-1}$) between them.
\cite{tecza00} find evidence for a recent starburst in the
central kpc (the region encompassing the two nuclei), and \cite{pol07} resolve a few dozen clusters in the
nuclear region that are also consistent with a
recent starburst.

In this paper we build on the qualitative picture put forward by \cite{tecza00} that
NGC\,6240 is a merger already past its first close encounter, where
the luminosity arises from a very recent burst of star formation, and therefore
in which the mass must be due to an older stellar population.
We use new adaptive optics integral field K-band
spectra, interferometric mm CO(2-1) data, and merger star formation histories from numerical simulations, to perform a quantitative analysis of the mass,
luminosity, age, and origin of the young and old stellar populations.

We first introduce the observations and data reduction processes for
the data sets used in our analyses (\S\ref{sec:obs}). We briefly look
at the molecular gas (\S\ref{sec:gas}), and then focus on a detailed
analysis of the stellar dynamics (\S\ref{sec:extinction},
\S\ref{sec:kin:BH}, \S\ref{sec:stellkin}, \S\ref{sec:kin:jeans}),
before discussing the merger geometry and stage, and the curious CO(2-1) morphology (\S\ref{sec:sims}). Bringing
together these results, we then investigate the scale of the starburst
(\S\ref{sec:starburst}) and nature and
origin of the nuclei (\S\ref{sec:SFhistory}). We summarise our results
in \S\ref{sec:sum}.

\section{Observations and Data Processing}
\label{sec:obs}

\subsection{SINFONI Data}

\subsubsection{Observations and Reduction}
\label{sec:obs:red}

Observations of NGC\,6240 were performed on the night of 20~Aug~2007
at the Very Large Telescope (VLT) with SINFONI.
SINFONI is a near-infrared integral field spectrometer \citep{eis03}
which includes a curvature-based adaptive optics system
\citep{bon04}. 
For these observations the wavefront reference was provided by the
Laser Guide Star Facility \citep{bon06,rab04}.
We used the $R=13.7$\,mag star 40\arcsec\ off-axis from NGC\,6240 to correct for tip-tilt motions.
The instrument was rotated by 70$^\circ$ east of north in order to
acquire this star. 
In order to cover both nuclei of NGC\,6240 simultaneously, the
0.05\arcsec$\times$0.10\arcsec\ pixel scale was 
chosen, giving a 3.2\arcsec$\times$3.2\arcsec\ field of view.
This pixel scale and the K-band grating provide a resolution
of $6.6\times10^{-4}$\,$\mu$m, equivalent to $R\sim3300$ or
90\,km\,s$^{-1}$ FWHM at 2.18\,$\mu$m.

The data set comprised 12 object frames interleaved with 4 sky
frames. 
Each exposure was 300\,seconds, yielding a total on-source integration
time of 60\,minutes. 
Data were reduced using the dedicated SPRED software package
\citep{abu06}, which provides similar processing to that for long-slit
data but with the added ability to reconstruct the data cube by
stacking the individual slitlets. 
In order to optimise the sky subtraction, we made use of the algorithm described in
\cite{dav07oh}.
The image was rotated to restore the standard orientation, with north
along the Y-axis, and east along the X-axis.
The pixel scale of the final reduced cube was $0.05\arcsec \times
0.05\arcsec$.

Standard star frames were similarly reconstructed into cubes. 
Telluric correction and flux calibration were performed using
HD\,147550, a B9\,V star with K=5.96\,mag.
Velocity Standards were also observed with the same spatial and
spectral settings, in order to enable recovery of the kinematics.
The 3 stars observed, and added to our library of similar stars, were
HD\,176617 (M3\,III), HD\,185318 (K5\,III), and HD\,168815 (K5\,II).
These stars were chosen because of their deep CO absorption longward
of 2.3\,$\mu$m which provides a good match to that of NGC\,6240.

\subsubsection{PSF Estimation}
\label{sec:obs:psf}

\begin{figure}
\begin{center}
\includegraphics[height=0.24\textheight]{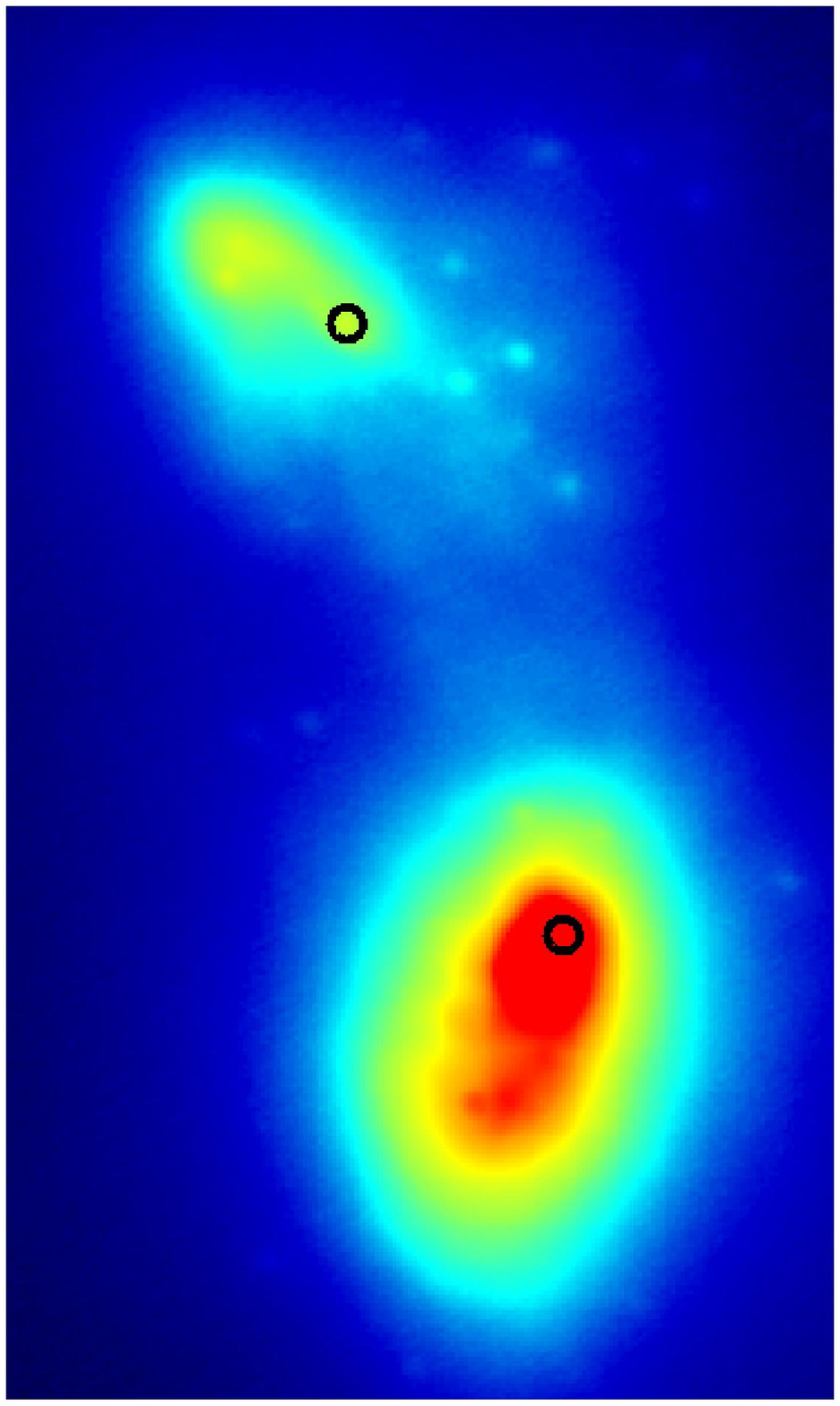}
\includegraphics[height=0.24\textheight]{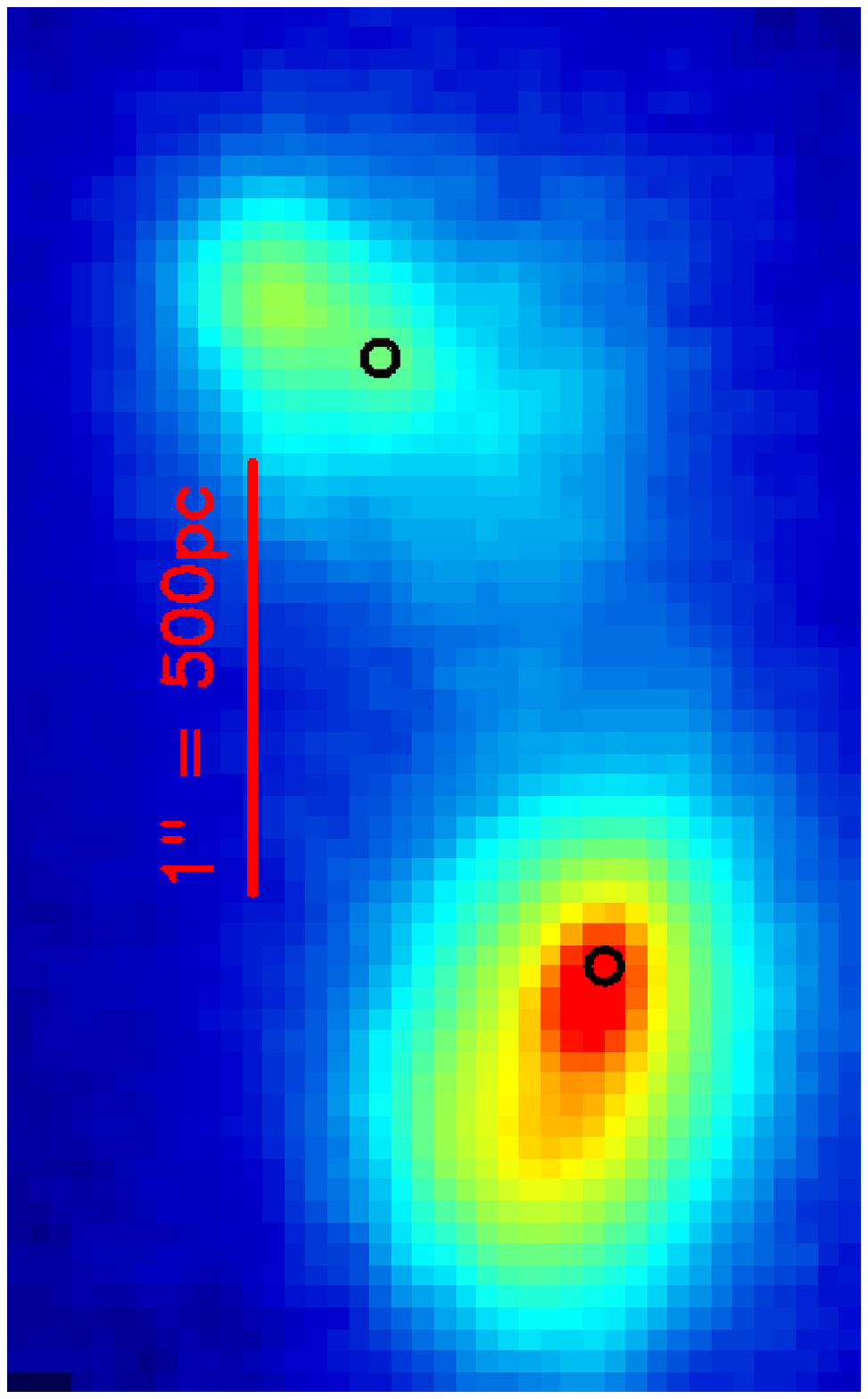}
\caption{K-band adaptive optics images of NGC\,6240 in square-root
  colour scale.
Left: K-band flux density from NIRC2 on Keck\,II
(pixel scale 0.01\arcsec, \citealt{pol07,max07}).
Right: 2.2\,$\mu$m continuum from SINFONI on the VLT (pixel scale 0.05\arcsec), covering
  approximately the same field as left. 
In both panels, circles indicate AGN positions (see
  \S\ref{sec:kin:BH}), north is up and east is left, 
and the red bar indicates 1\,\arcsec\ (500\,pc).}
\label{fig:BHpositions}
\end{center}
\end{figure}

We estimate the PSF by comparing the SINFONI data to higher
resolution K-band Keck AO imaging data \citep{pol07}, which are both shown in
Fig.~\ref{fig:BHpositions}.
Although the spatial resolution of our integral field observations is not as good as that achieved by these Keck imaging data, we note that the two bright spots in the northern nucleus are
clearly separated in our data, indicating that our resolution is
better than the 0.2\arcsec\ achieved by HST/NICMOS (Fig.~3c in
\citealt{gerssen04}).

As outlined by \cite{dav08} and employed by \cite{mue06}, 
since an observed image is the convolution of an intrinsic
image with a PSF, $I_{obs} = I_{intr} \otimes PSF$, we can use
a higher resolution image with a known PSF to obtain a good estimate
of the PSF in a lower resolution image in two steps.
Having resampled the data to the same pixel scale, we first
find the broadening function $B$ whose convolution with the high
resolution image yields the best match to the low resolution image,
$I_{low} = I_{high} \otimes B$. 
Since the underlying intrinsic
image is the same for both $I_{low}$ and $I_{high}$, we can then
estimate the PSF of the low resolution image by convolving the
broadening function $B$ with the PSF of the high resolution image,
$PSF_{low} = PSF_{high} \otimes B$.
Since in this case $B$ dominates the size and shape of our PSF,
uncertainties in $PSF_{high}$ have little impact on $PSF_{low}$.
For the analyses carried out
here, this procedure gives a sufficiently accurate definition for the PSF. 
We find
the PSF to be well matched by an asymmetric Gaussian with major axis
oriented $-20^\circ$ east of north. The resolution is 0.097\arcsec$\times$0.162\arcsec\ FWHM, corresponding to $50\times80$\,pc at the distance of
NGC\,6240.
The asymmetry is to be expected since the resolution of the original data is
pixel limited, being sampled at 0.05\arcsec$\times$0.10\arcsec.
The reason we are able to do better than the formal Nyquist limit is because the SINFONI data are resampled to 0.05\arcsec$\times$0.05\arcsec and many dithered exposures are combined using sub-pixel shifts to align them.

\subsubsection{Spatial binning}
\label{sec:obs:binning}

Because of the limited signal-to-noise in our data, we have binned
them spatially using an optimal Voronoi tessellation \citep{cap03}.
This algorithm bins pixels together into groups by accreting new pixels
to each group according to how close they are to the centroid of the
current group. 
The resulting groups then provide a set of positions (centroids) and
mean fluxes which are used as the initial `generators'.
A further algorithm optimises the generator
configuration based on a centroidal Voronoi tessellation.
The final set of generators are the positions of the flux-weighted
centroids of each binned group, and have the property that each pixel
in the original image is assigned to the group
corresponding to the nearest generator.
This procedure only affects spaxels below a specified signal-to-noise
threshold, and hence does not impact the spatial resolution of high
signal-to-noise regions. 
The signal-to-noise cutoff was 20, chosen such that the regions with 
the highest signal-to-noise remained fully sampled, but lower
signal-to-noise regions were binned. 
This avoided compromising the spatial resolution around the center of
each nucleus, while enabling us to extend the region in which analyses
can be performed.
The effect of this can be seen in
the resulting stellar kinematics maps, where the colours in the outer
regions appear in blocks rather than individual pixels.
The binning scheme that this routine provided was applied to each
spectral plane in the cube, and the kinematics were extracted from
each bin as described below.

\subsubsection{Extracting Stellar Kinematics}
\label{sec:obs:stellar}

\begin{figure}
\includegraphics[width=0.48\textwidth]{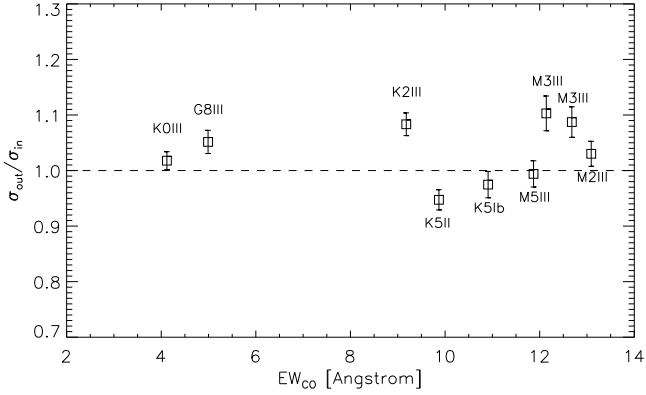}
\caption{The ratio between the measured dispersion $\sigma_{\rm out}$
  and the input dispersion $\sigma_{\rm in}$ for a range of different
  stellar templates used to extract kinematics from a K\,5Ib template.
For each combination, a variety of dispersions in the range
  150--300\,km\,s$^{-1}$ were used and noise was added to achieve a
  S/N$\sim$35.
Variations in measured stellar kinematics using templates with a large range of
  W$_{CO}$ and stellar types are around a few percent, showing that
  with our method a mismatched stellar template has little effect.
}
\label{fig:sigmas}
\end{figure}

\begin{figure}
\includegraphics[width=0.48\textwidth]{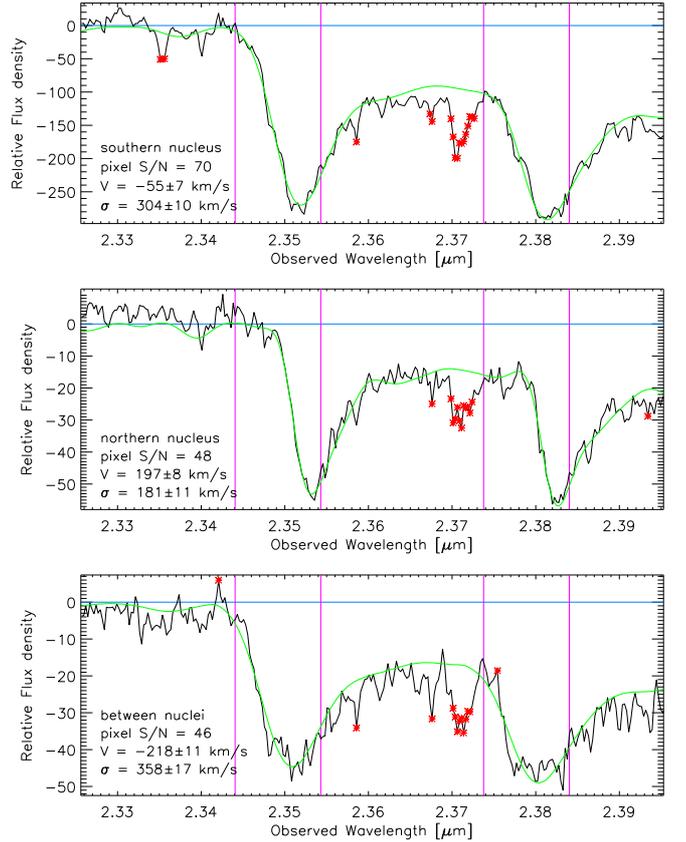}
\caption{Spectra (black lines) and template fits (green lines) to the
  CO\,2-0 and CO\,3-1 stellar absorption features.
These are for individual bins in the southern nucleus (top), the northern nucleus (middle)
  and in the region of high dispersion between them (bottom).
 The blue line indicates the continuum (i.e. zero) level;
  red asterisks denote pixels rejected during the fit;
  the pairs of magenta lines enclose regions around the bandheads for
  which the weighting is enhanced during the fit.
These plots demonstrate the quality of the fit using our single
  M3\,III template.
}
\label{fig:bandheads}
\end{figure}

The observed wavelength regime contains the stellar CO absorption bandheads
longward of 2.29\,$\mu$m, whose sharp blue edges are very sensitive to stellar
motions.
We utilise CO\,2-0 and CO\,3-1 to derive 2D~maps of the stellar velocity and
dispersion.
The spectra were prepared by normalising them with respect to a linear
fit to the line-free continuum.
The continuum level was then set to zero by subtracting unity.
A template spectrum, that of the M3\,III star HD\,176617, was prepared in the same way.
Although both \cite{sug97} and \cite{tecza00} used K\,Ib stars, they
also both noted that M\,III stars provided almost equally good fits to
the K-band features (see Fig.~5 in \citealt{tecza00}).
We therefore do not expect template mismatch to be a problem but, as
explained below, our method furthermore minimises the impact of any
discrepancy.
\cite{sil03} showed that the dispersion measured depends on the equivalent width of
the template; this can be understood since the depth of the extinction feature relative
to the continuum level (which may contain non-stellar emission)
places a strong constraint on the fitting parameters. But it is the
spectral width, not the equivalent width (i.e. depth),
of the absorption feature that is important for deriving the dispersion. 
In our fitting method we resolve the problem of the feature's depth by
setting the continuum level 
to zero. Therefore matching the relative depth of the absorption features, and
hence the choice of template, is no longer critical.

\begin{figure*}
\begin{center}
\includegraphics[width=0.2\textwidth]{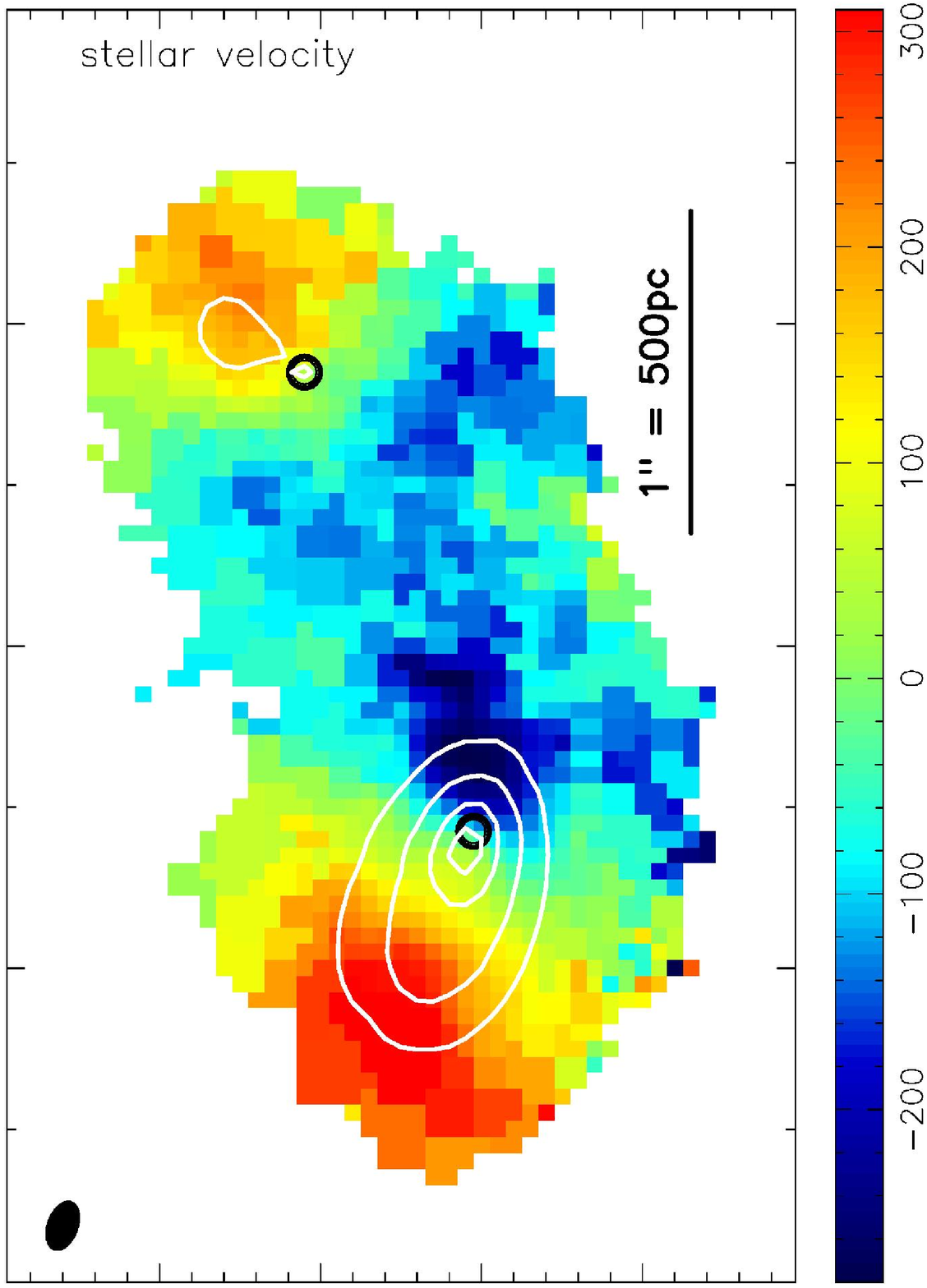}
\hspace{5mm}
\includegraphics[width=0.2\textwidth]{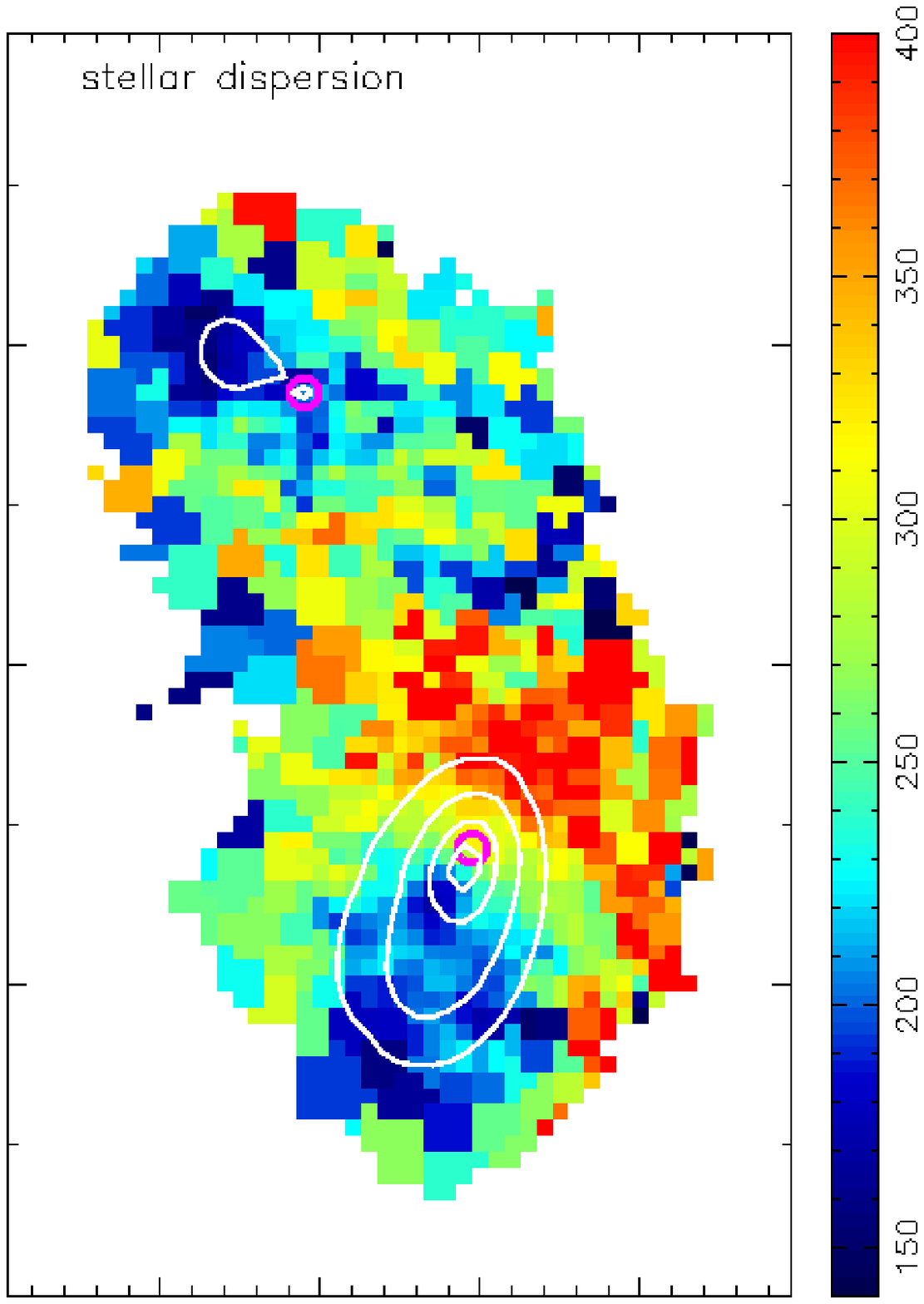}
\hspace{5mm}
\includegraphics[width=0.2\textwidth]{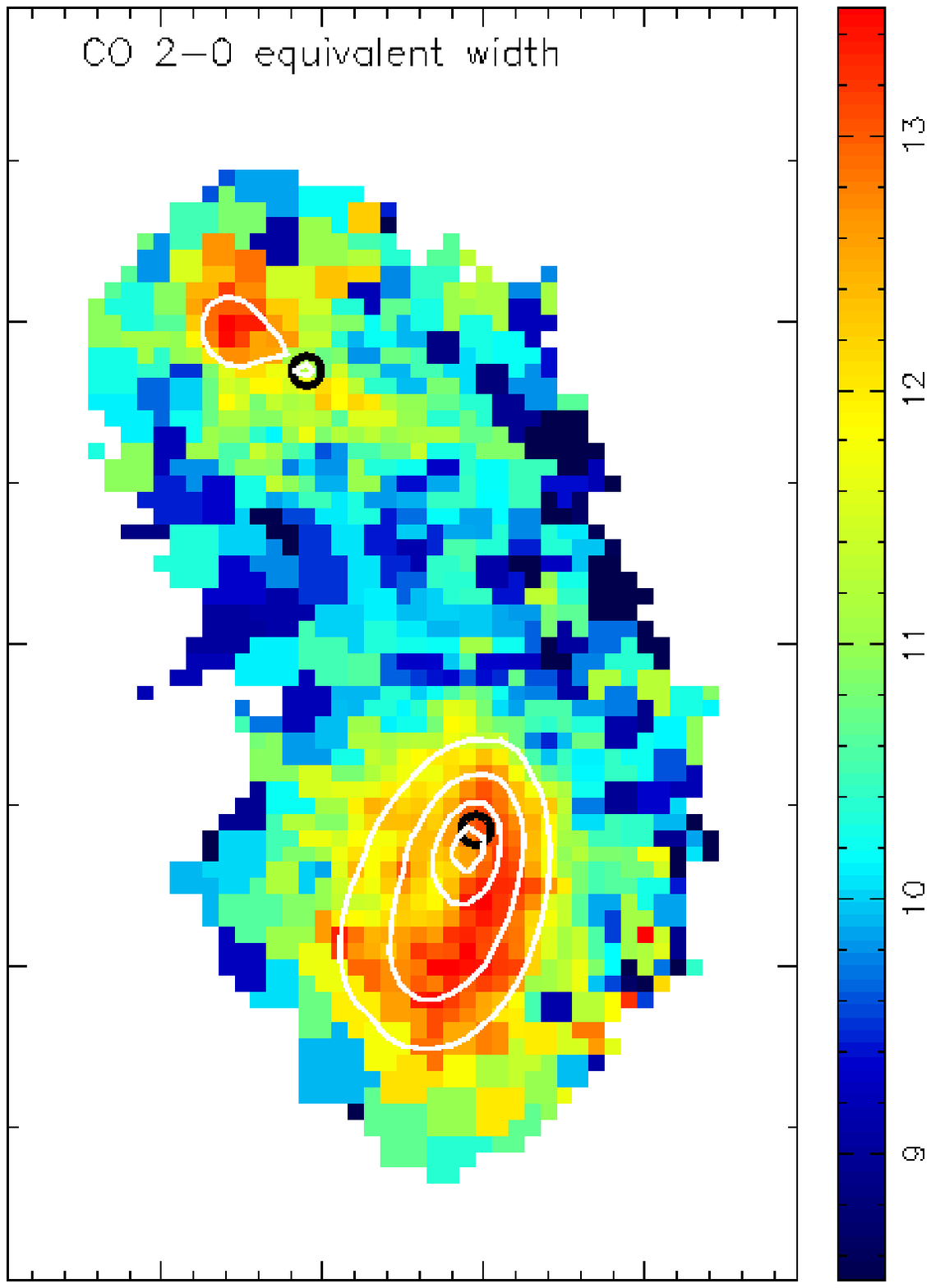}
\end{center}
\begin{center}
\includegraphics[width=0.2\textwidth]{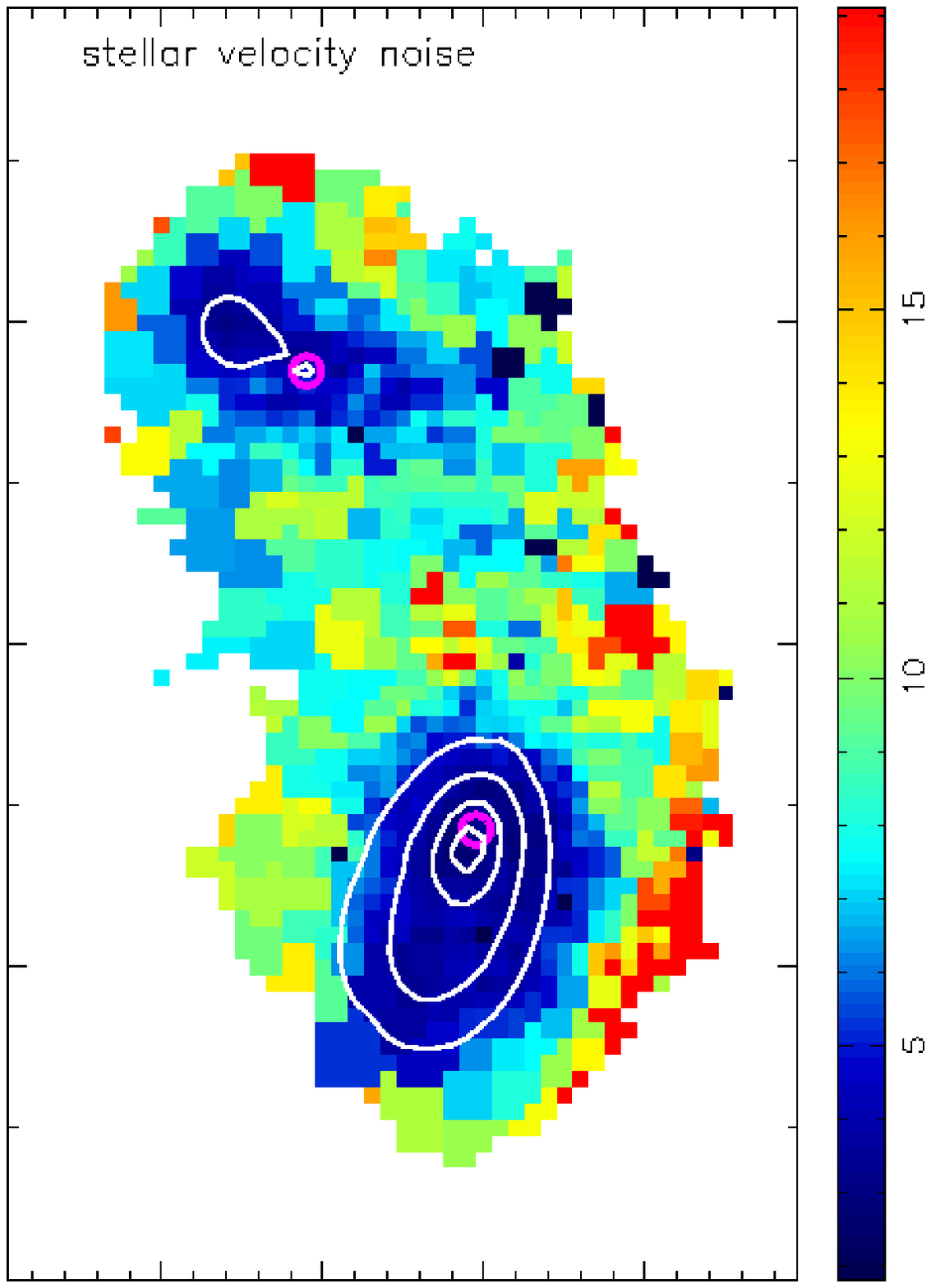}
\hspace{5mm}
\includegraphics[width=0.2\textwidth]{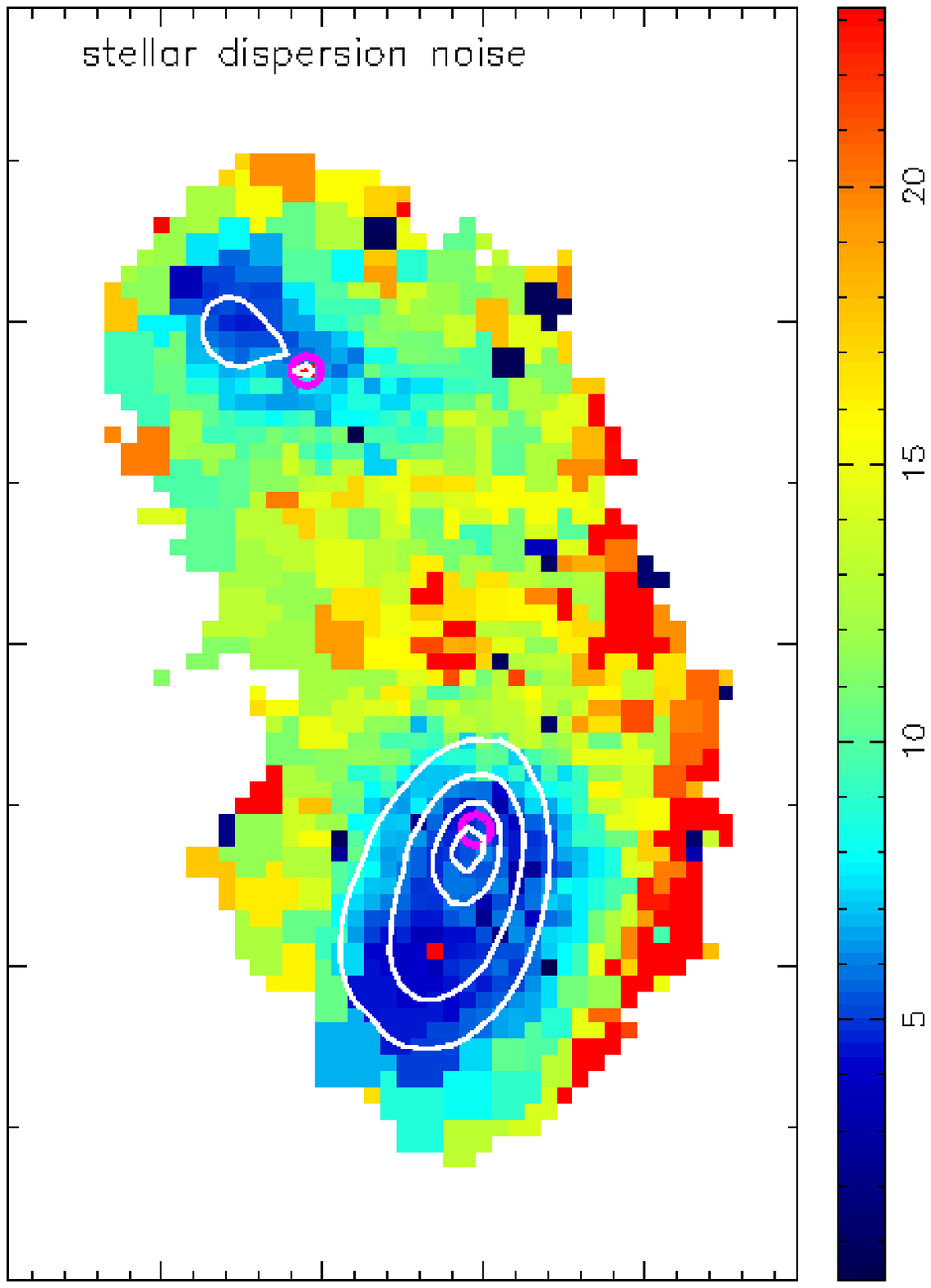}
\hspace{5mm}
\includegraphics[width=0.2\textwidth]{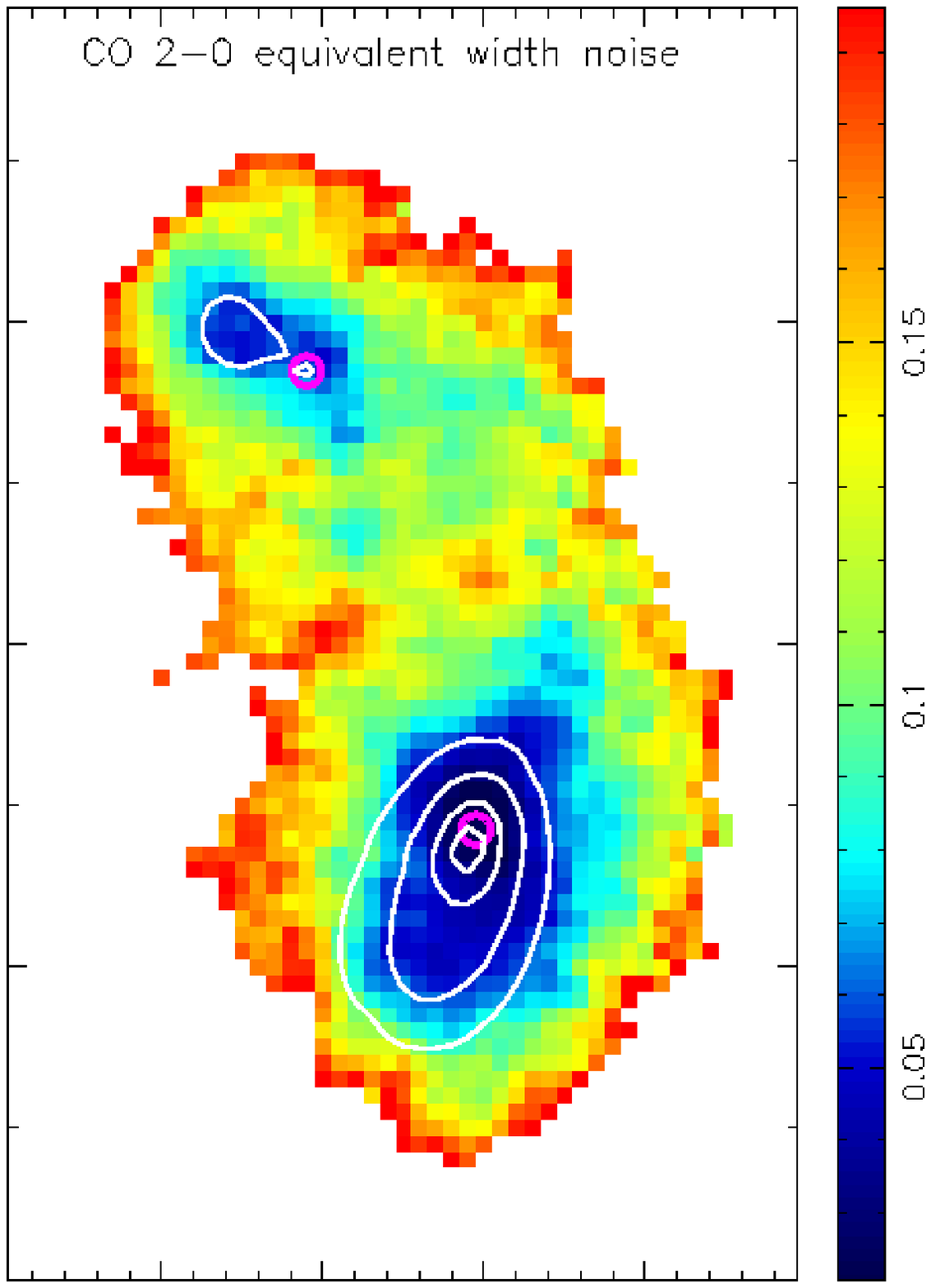}
\end{center}
\caption{Stellar velocity and corresponding noise map (top and bottom left, both in km\,s$^{-1}$ with respect to systemic), dispersion and corresponding noise map (middle, both in km\,s$^{-1}$),
and $W_{CO\,2-0}$ and its noise map (right, both in \AA).
Contours trace the continuum as in Fig.~\ref{fig:BHpositions}.
Black/magenta circles mark the AGN positions (see \S\ref{sec:kin:BH}).
The bar indicates 1\arcsec\ (images are 2.45\arcsec$\times$3.95\arcsec), the PSF size is shown in the lower left corner of panel 1. North is up and east is to the left.}
\label{fig:stellar}
\end{figure*}

In order to demonstrate this quantitatively, we have made use of the stellar template library of \cite{winge09}.
We created a large series of mock observations by broadening a K5\,Ib
template by various amounts in the range 150--300\,km\,s$^{-1}$, and
adding noise.
We then extracted the dispersion using various
mis-matched stellar templates spanning a
wide range of W$_{CO}$ (equivalent width of the CO absorption feature)
and spectral type. 
As can be seen in Fig.~\ref{fig:sigmas}, our method recovers the input
dispersion with a typical error of only a few percent, regardless of
the template's spectral type.
This shows that our method of extracting the stellar kinematics works
reliably even if the template used is not a perfect fit. 
We emphasize that, in contrast to \cite{sug97} and \cite{tecza00}, we
do not use the quality of the template fit to draw conclusions about the
prevalent stellar population. This requires caution, and we
discuss the pitfalls associated with doing so in \S\ref{sec:starburst}.
As such, because we use this star only for extracting kinematics, our
choice of template can be fully justified by the good fits to the
bandheads in the galaxy spectra in Fig.~\ref{fig:bandheads}.

For the spectrum at each spatial pixel (`spaxel' hereafter) in the data cube, we convolve the
template with a Gaussian, adjusting the properties of the Gaussian to
minimise the difference between the convolution product and the galaxy
spectrum.
Deviant pixels are rejected from the fit.
The spectral regions covering the steep edges of the bandhead at all
expected stellar velocities are given five-fold weight
(Figure~\ref{fig:bandheads}), in order to focus the fit on the
kinematics rather than details in the spectrum.
To ensure consistency in our analyses, the spectral ranges applied to
compute W$_{CO}$ were taken from \cite{for00}, since these ranges are
also used by the stellar synthesis code STARS which we use later.
A high signal-to-noise ($\gtrsim50$, \citealt{cap09}) is required in
order to measure deviations in the line of sight velocity distribution
from a simple Gaussian.
Without binning beyond a useful spatial resolution, the
signal-to-noise in most of our spectra is not sufficient to include
the Gauss-Hermite terms $h_3$ and $h_4$ in the fit.
We therefore limited the kinematic extraction to $V$ and $\sigma$.
The noise in the kinematics was determined using Monte Carlo
techniques: 100 realisations of the data were generated, a number that
ensures a fractional uncertainty in the noise of less than 10\%, by
perturbing the flux at each pixel according to its RMS.
These were fit using the same procedure as above.
Although the RMS at each pixel can be derived when combining
the data cubes, calculating the noise in bins is extremely difficult
due to the correlations between neighbouring pixels.
This is discussed in detail by \cite{for09}.
Instead, we have derived the noise as the RMS of the difference
between a spectrum and the convolved template (excluding rejected
pixels).
This may slightly over-estimate the noise, but will then yield
conservative error estimates on the kinematic parameters.
The resulting final maps of stellar velocity, dispersion, and
$W_{CO\,2-0}$ are displayed in Fig.~\ref{fig:stellar}.

\subsection{Plateau de Bure Interferometer Data}
\label{sec:bure}

\begin{figure*}
\begin{center}
\includegraphics[width=0.6\textwidth]{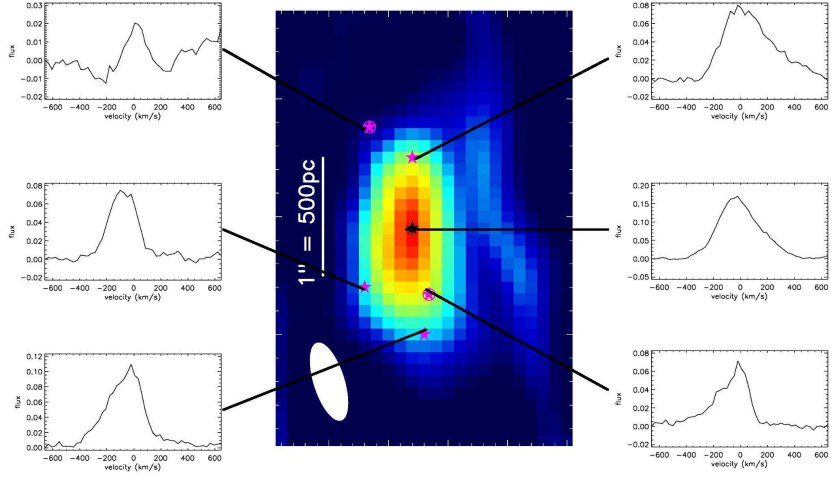}
\caption{Map of CO(2-1) flux with spectral line profiles shown for
  selected spaxels. Circles denote AGN positions, stars denote
  positions where spectra were extracted. The beam size is displayed
  in lower left.
The bar denotes 1\arcsec, and north is up and east is to the left.}
\label{fig:co}
\end{center}
\end{figure*}

\begin{figure*}
\begin{center}
\includegraphics[width=0.6\textwidth]{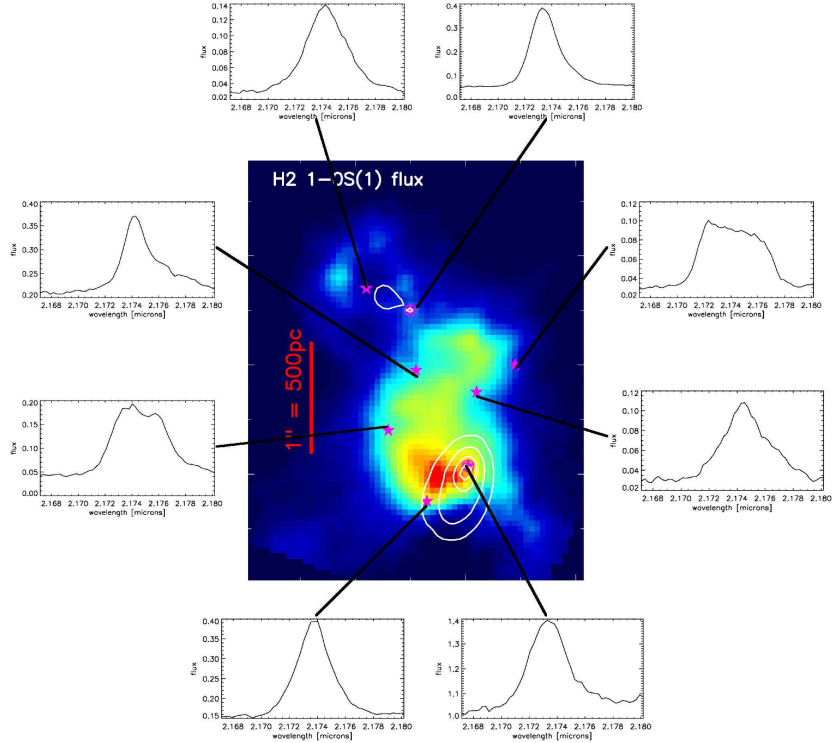}
\caption{Map of H$_2$ 1-0 S(1) emission, with line shapes at selected spaxels. Circles mark AGN positions, stars denote positions where spectra were extracted, white contours represent continuum emission.
The bar denotes 1\arcsec, and north is up and east is to the left.}
\label{fig:h2}
\end{center}
\end{figure*}

We have mapped the $^{12}$CO(J\,=\,2-1) line in NGC\,6240 with the IRAM millimetre interferometer, which is located at an altitude of 2550\,m on the Plateau de Bure, France \citep{guilloteau92}. The data were obtained in May 2007. The array consisted of six 15\,m antennae positioned in two configurations providing a large number of baselines ranging from 32 to 760\,m. We observed NGC\,6240 for 5\,hrs in A configuration, which provides a nominal spatial resolution of 0.35\arcsec\ at 230\,GHz. However, the low (+2deg) declination of NGC\,6240 limited the resolution in an approximately north-south direction. A spectral resolution of 2.5\,MHz, corresponding to 3.3\,km\,s$^{-1}$ for the CO(2-1) line, was provided by 8 correlator spectrometers covering the total receiver bandwidth of 1000\,MHz (1300\,km\,s$^{-1}$). All the data were first calibrated using the IRAM CLIC software. We then made uniformly weighted channel maps for the CO(2-1) data. We CLEANed all the maps using software available as part of the GILDAS package. To increase the sensitivity, maps were made with a velocity resolution of 26.6\,km\,s$^{-1}$. The CLEANed maps were reconvolved with a 0.69\arcsec$\times$0.26\arcsec\ FWHM Gaussian beam. The rms noise after CLEANing is 1.4\,mJy\,beam$^{-1}$.
Fig.~\ref{fig:co} shows the velocity-integrated CO(2-1) emission (black circles denoting approximate AGN locations), velocity moment map, and the dispersion (obtained from fitting Gaussians to the line shape).
The AGN positions were estimated using the positions given by \cite{gallimore04}; due to the astrometric uncertainties, these are only accurate to within 1-2 pixels (0.7-0.14\arcsec).

\section{Molecular Gas Emission}
\label{sec:gas}

\subsection{H$_2$ 1-0S(1)}
\label{sec:gas:h2}

\begin{figure}
\begin{center}
\includegraphics[width=0.24\textwidth]{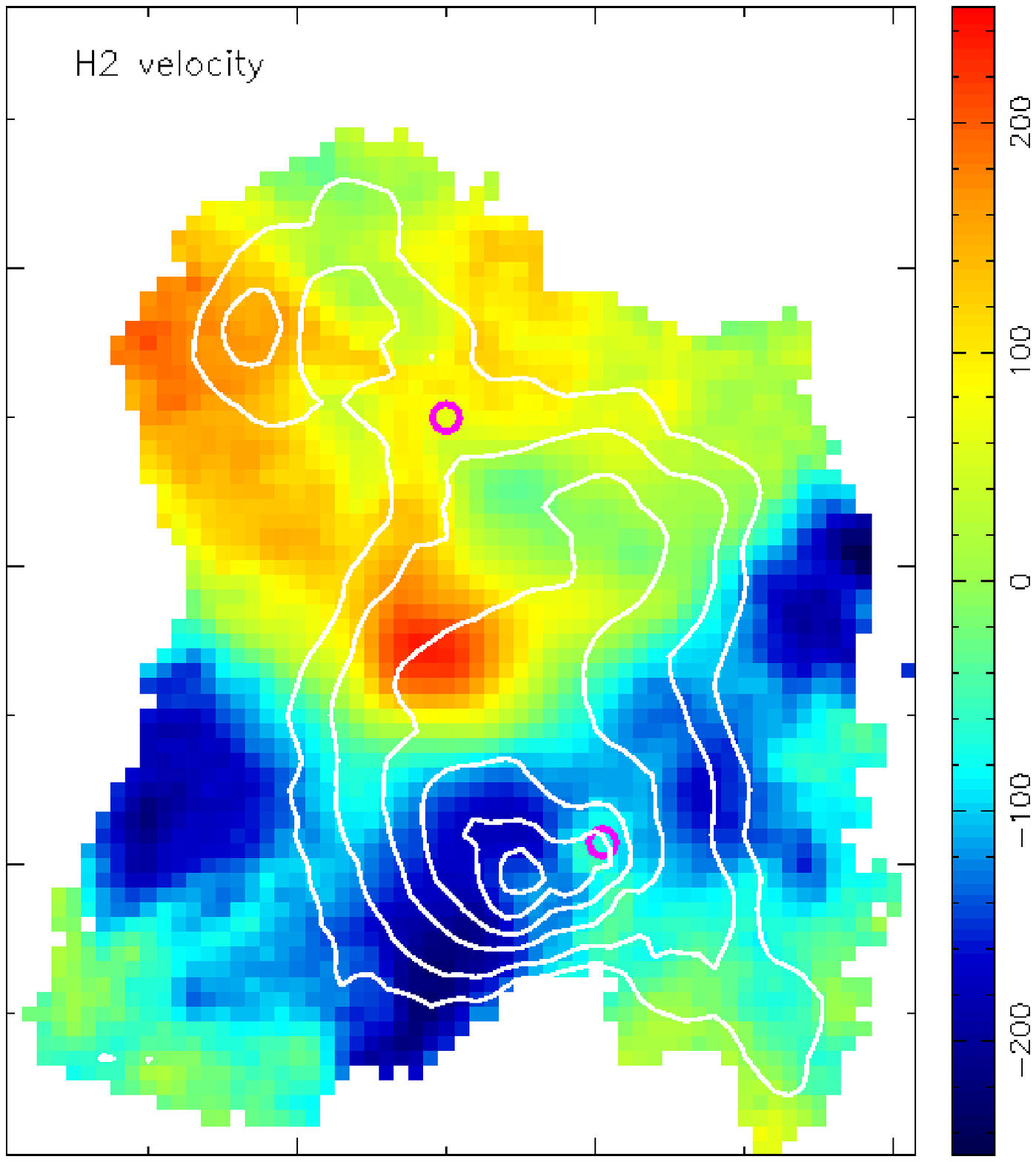}
\includegraphics[width=0.24\textwidth]{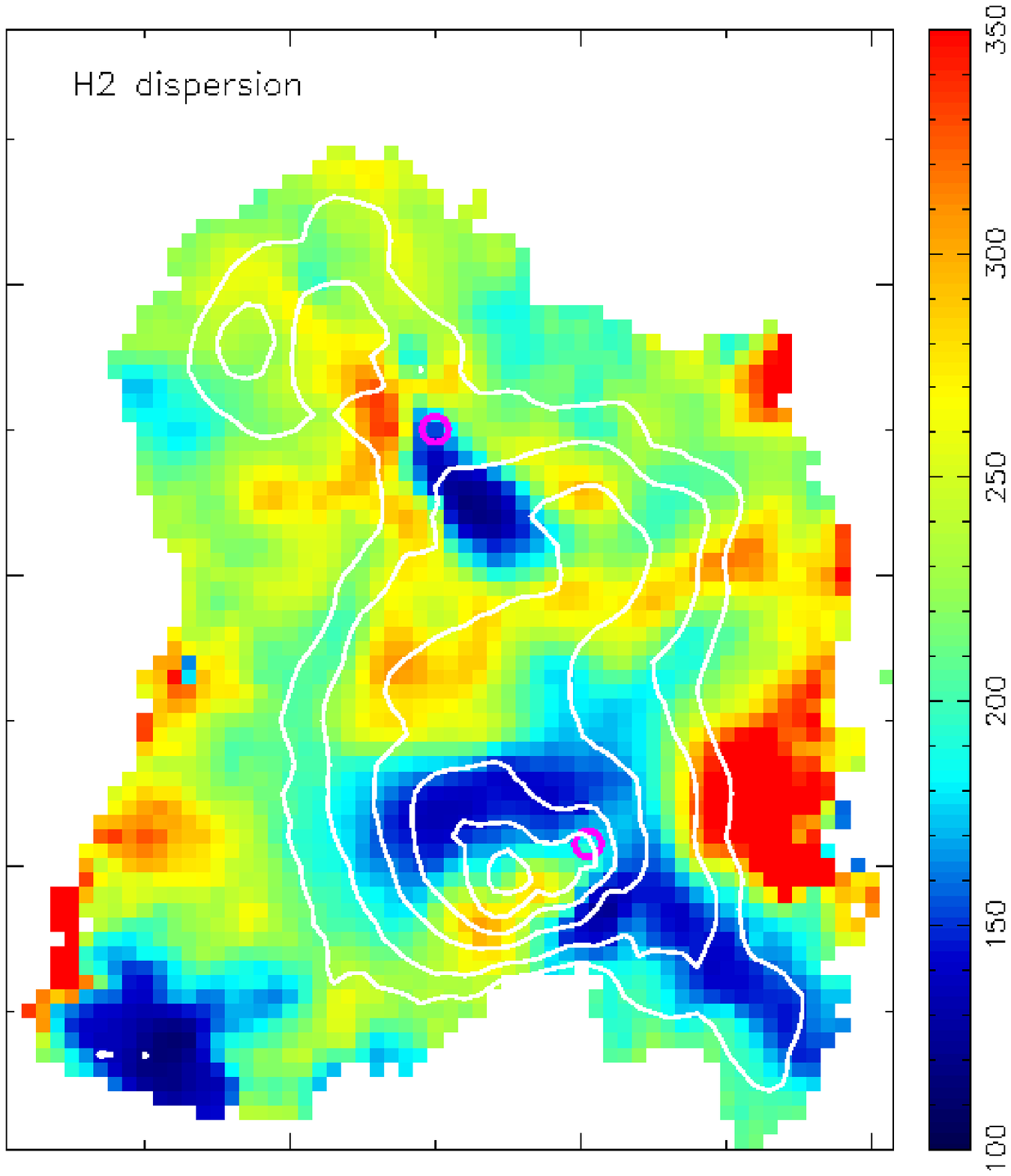}
\caption{Velocity moment (left) and dispersion (right) maps of H$_2$. Magenta circles mark AGN positions, white contours represent H$_2$\,1-0\,S(1) line emission.}
\label{fig:h2b}
\end{center}
\end{figure}

NGC\,6240 has the strongest H$_2$ line emission found in any galaxy to
date \citep{joseph84}. Several authors \citep{werf93,sug97,tecza00}
identify shocks as the excitation mechanism, and \cite{ohy03} use the
relative intensity of H$_2$ and CO(2-1) emission to conjecture that
the shocks occur due to a superwind outflowing from the southern
nucleus colliding with the molecular gas concentration. Our new
SINFONI data provide the most detailed spatial and spectral view 
currently available. As can be seen in Fig.~\ref{fig:h2}, the line
profiles are highly complex, with clear evidence of multiple
components. This, and the dispersion map (Fig.~\ref{fig:h2b}), support
the picture of highly disturbed, turbulent gas. The emission
morphology is uncorrelated with the stellar emission distribution, but
roughly consistent with the cold gas emission (as traced by the
CO(2-1) emission, \S\ref{sec:gas:CO}). The velocity moment map
(Fig.~\ref{fig:h2b}) shows a global velocity gradient, again broadly
consistent with what is seen in CO(2-1).

\subsection{CO(2-1)}
\label{sec:gas:CO}

Fig.~\ref{fig:co} shows the velocity-integrated CO(2-1) emission, with
line shapes at selected positions.
With a factor of two improvement in resolution along the E-W
direction, our data re-affirm the findings of \cite{tac99}.
However, our resolution in the crucial N-S direction is limited by
the declination of the source. As a result we are not able to draw
more detailed conclusions about the CO(2-1).

The CO emission is concentrated in between the two nuclei, and the
CO(2-1) velocity map, red and blue wings, and P-V-diagram all display
signs of a velocity gradient. This is consistent with the findings of
\cite{tac99}, who conclude that the molecular gas likely is
concentrated in a self-gravitating rotationally supported, but highly
turbulent disc in the internuclear region. 
However, such a central gas
concentration is not expected from merger simulations, which generally
predict the gas to remain largely bound to the progenitors until the
nuclei coalesce. We discuss this issue further in \S\ref{sec:sims}.

The total flux is about 40\% less than that measured by \cite{tac99},
indicating that our smaller beam has resolved out some emission on
intermediate to large scales.
We have not made a correction for this, since we are interested
primarily in the small scale emission of the nuclei themselves, where
the effect is likely to be negligible.
To estimate the gas mass associated with the nuclei, we have measured
velocity-integrated line fluxes within the same radii as those used in
our dynamical mass modelling (\S\ref{sec:kin:jeans}), which 
were chosen to cover the extent of observed stellar rotation in the
nuclei.
In the northern nucleus we find 
$S_{CO}\Delta V = 37$\,Jy\,km\,s$^{-1}$ out to a radius of 250\,pc;
in the southern nucleus we find 
$S_{CO}\Delta V = 178$\,Jy\,km\,s$^{-1}$ out to a radius of 320\,pc.

We estimate the gas mass by converting to line luminosity
L'$_{CO}=3.25\times10^7S_{CO}\Delta V \nu_{obs}^{-2}D_L^2(1+z)^{-3}$
(with $D_L$=\,97\,Mpc, z\,=\,0.0243,
$\nu_{obs,CO_{2-1}}$=\,225.1\,GHz), and using the relations 
L(CO$_{2-1}$)/L(CO$_{1-0}$)\,=\,0.8 \citep{casoli92} and
M$_{H_2}$/L$_{CO_{1-0}}\sim$\,1\,M$_{\odot}$/(K\,km\,s$^{-1}$\,pc$^2$),
as \cite{downessolomon98} find for ULIRGs.
This yields $\sim0.2\times10^9$\,M$_\odot$ and
$\sim1.1\times10^9$\,M$_\odot$ for 
northern and southern nuclei, respectively.
Similarly, for an aperture of 1\arcsec\ diameter centred on the CO
emission peak, we derive a gas mass of $\sim$3.1$\times10^9$\,M$_\odot$.
We caution that the nuclear masses are very uncertain and should be
treated only as order-of-magnitude estimates.
The reason is that a significant fraction of the gas within the apertures may not
be physically co-located with the nuclei, which would imply the masses
are upper limits.
On the other hand, the CO abundance around the nuclei is very likely
to have been significantly reduced by X-ray irradiation from the AGN
(see \S\ref{sec:sims}).
In this case, the masses would be underestimates.

\begin{figure}
\begin{center}
\includegraphics[width=0.20\textwidth]{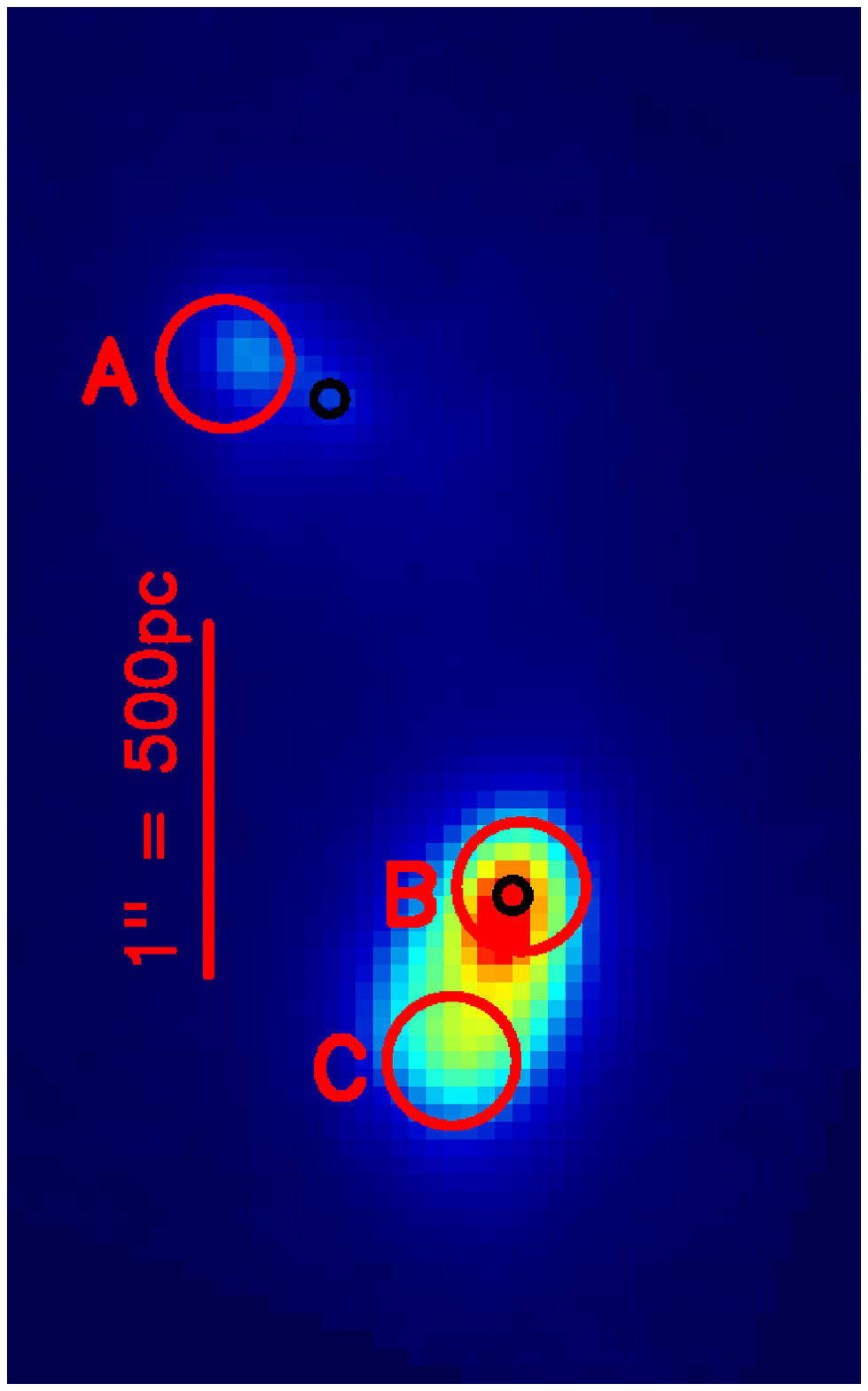}
\includegraphics[width=0.28\textwidth]{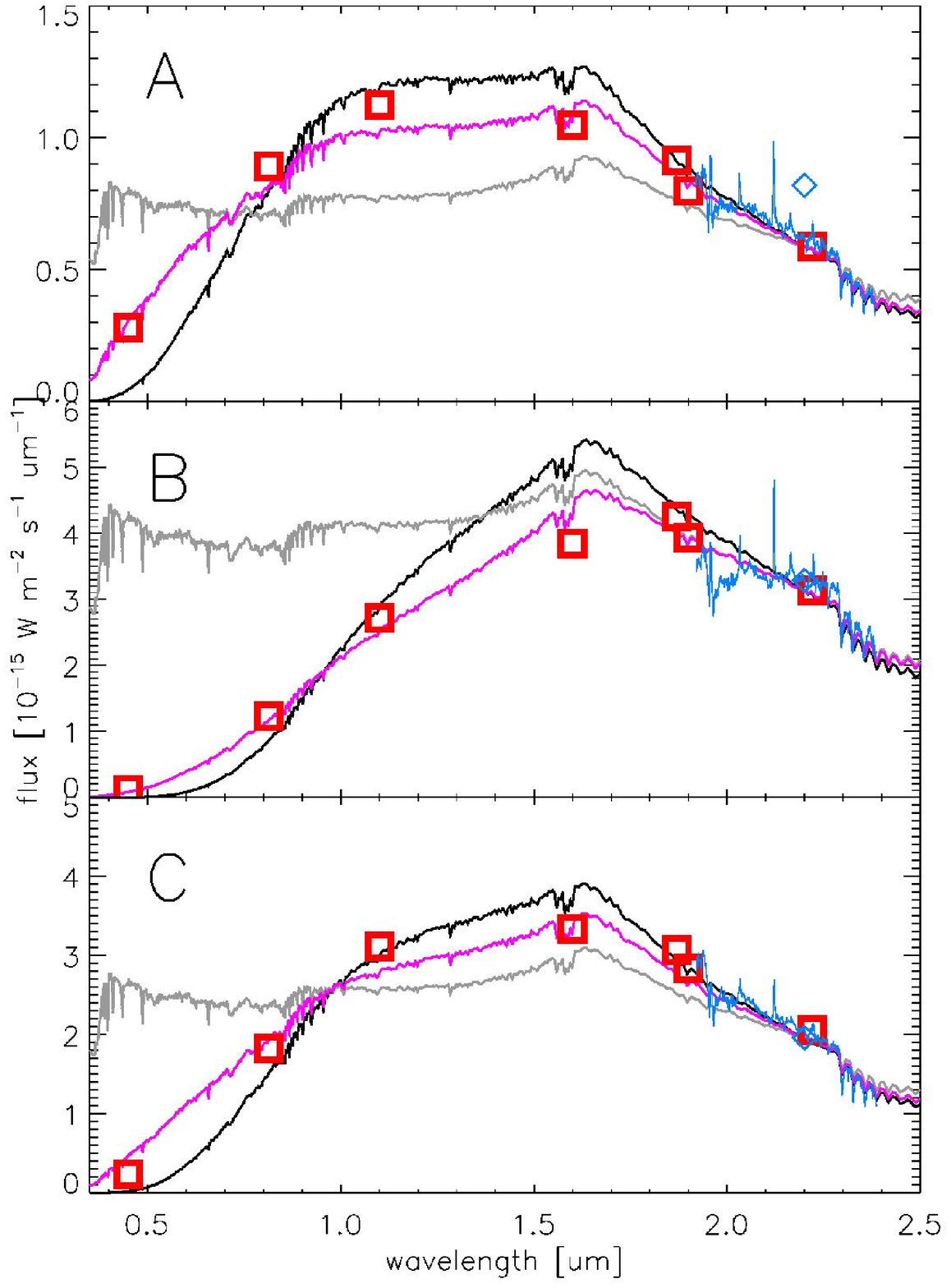}
\caption{Comparison of different extinction models: We measured HST photometric data points (red squares) in 0.2\arcsec\,radius apertures; positions for the three examples shown here are indicated in the left panel. To these we fitted a synthetic STARS spectrum which was reddened using different reddening prescriptions: screen extinction (black), mixed extinction (grey), and the \cite{calzetti00} reddening law (magenta). We also overplot our SINFONI spectrum (blue), scaled to the HST 2.22$mu$m data point (absolute SINFONI flux at 2.2$mu$m shown as blue diamond). As can be seen, the \cite{calzetti00} reddening law provides the best fit to the data.}
\label{fig:extinctionspectrum}
\end{center}
\end{figure}

\section{Extinction and Luminosity of Nuclei}
\label{sec:extinction}

Like most ULIRGs, NGC\,6240 contains significant amounts of dust
\citep{tecza00}, and hence correcting any flux measurement for
extinction is paramount to ensure accuracy in the analyses described
below. It is valuable to consider the extinction derived from mid-infrared
data to assess whether the near infrared might be affected by
saturation. 
However, this appears to be uncertain -- primarily because no 
H{\sc ii} lines were detected either by ISO \citep{lut03} or Spitzer
\citep{arm06}.
Based on silicate absorption at 9.7\,$\mu$m, \cite{arm06} find an
extinction of $A_V$\,$\sim$\,95\,mag to the coronal line region.
The more moderate estimates presented by \cite{lut03} are for a global
extinction, and correspond to a dust screen $A_V$\,$\sim$\,15--20\,mag.
This estimate is comparable to derivations at 
radio ($N_H\sim$(1.5--2)$\times10^{22}$\,cm$^{-2}$, \citealt{bes01})
and X-ray ($N_H\sim10^{22}$\,cm$^{-2}$, \citealt{kom03}) wavelengths. 
This is important, because it implies a modest typical K-band
screen extinction of only $A_K$\,$\sim$\,1--2\,mag.
Thus our data
should be sensitive to the majority of the emission.
However, we note that the X-ray spectrum at energies above 10\,keV
indicates that at least one of the AGN themselves may be obscured by a
column $>10^{24}$\,cm$^{-2}$ \citep{vig99,ike00,net05}. Differences
between these various column density measurements are to be expected
since they sample different spatial scales and sight lines; for example,
the X-ray data specifically measure gas column density
along the line of sight to the AGN rather than the extended cold dusty
medium.

When correcting for the obscuring effect of dust, one needs to make
assumptions about its location and distribution.
The two most commonly used models assume either a uniform dust screen between the observer
and the stars, leading to a reduction in observed flux according to
$F_{obs}/F_{em} = e^{-\tau}$. Or the stars and dust are assumed to be spatially
coincident and uniformly mixed (`mixed model'), in this case the observed flux decreases as
$F_{obs}/F_{em} = (1 - e^{-\tau})/\tau$. 
Another alternative is the \cite{calzetti00} reddening
law, derived empirically from observations of starburst galaxies.
Its wavelength dependence reflects both the dust grain properties and
the distribution of the dust with respect to the
stars in these galaxies. 
The effect of this reddening
law is remarkably similar to a combination of mixed and screen
extinction, with -- for the degree of extinction seen in NGC\,6240 --
the mixed component dominating in the infrared, and the screen
component increasingly important at optical wavelengths.
In order to investigate which extinction model best captures the characteristics of the dust
distribution of NGC\,6240, we obtained a number of archival HST imaging data, spanning wavelengths
from 0.45$\mu$m to 2.22$\mu$m, and extracted photometric data
points from a number of 0.2\arcsec\,radius apertures across our field
of view. 
And we calculated a synthetic stellar spectrum using the
stellar synthesis code STARS
\citep{sternberg98,ste03,for03,dav03,dav05,dav06,dav07}, assuming a
star formation rate typical for a merger (`Antennae'-simulation,
\S\ref{sec:sims}). We then fitted this spectrum to the HST data points
and our SINFONI K-band spectra, using the screen and mixed extinction
models, and the \cite{calzetti00} reddening law. As can be seen in
Fig.~\ref{fig:extinctionspectrum}, the \cite{calzetti00} reddening law
best reproduces the observations;
the mixed model saturates at optical
wavelengths and cannot redden the spectra sufficiently, and the screen
model is reddening the spectra too much at shorter
wavelengths. 
We also test whether our choice of star formation history has an
influence on this result, by conducting the same test with synthesised
spectra for a 20\,Myr old instantaneous starburst and 1\,Gyr of
continuous star formation.
Both also indicate the \cite{calzetti00} reddening law to be appropriate.

\begin{figure}
\begin{center}
\includegraphics[width=0.25\textwidth]{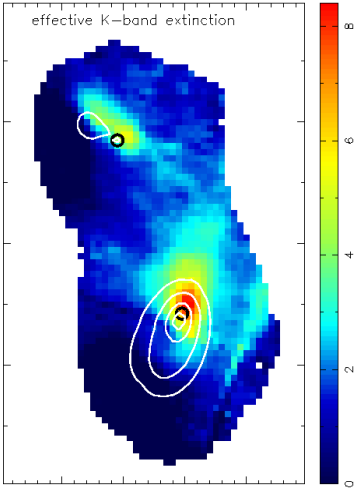}
\caption{Effective K-band extinction A$_K$.
As for Fig.~\ref{fig:BHpositions}, contours trace the continuum and
  black circles denote the AGN positions.
Each 0.05\arcsec\ pixel length corresponds to 25\,pc at the distance of NGC\,6240.}
\label{fig:extinction}
\end{center}
\end{figure}

We can then find $F_{obs}/F_{em}$ for each spatial pixel by adjusting
the reddening for a set of stellar template spectra according to
Eqn.~2 in \cite{calzetti00}, until the best-fit of a linear
combination of templates to the measured line-free continuum is
achieved. 
Fig.~\ref{fig:extinction} shows the resulting map of $A_K$.
The choice of stellar templates does not affect the result, since the K-band samples the Rayleigh-Jeans tail of the blackbody curve and hence to a good approximation all stars, and late-type stars in particular, have the
same spectral slope.

One might ask whether a component of non-stellar emission from hot
dust might be present, in our field of view in general and close to
the AGN in particular. This is an important question, since it would
have an influence on the stellar masses and mass-to-light ratios we
calculate later on.
That the K-band continuum may even be dominated by hot dust has been
proposed by \cite{arm06}.
However, this conclusion was based on a fit to the near-IR spectral energy distribution that was constrained
primarily as the residual in the blue side of a much more dominant
cooler component, and is therefore rather uncertain.
Instead, the good match of reddened synthetic stellar spectra to the
photometry (Fig.~\ref{fig:extinctionspectrum}) indicates that
non-stellar emission is unlikely to contribute significantly. 
More importantly, emission from hot dust would dilute the stellar CO
absorption features. Since we measure
$W_{CO\,2-0}$\,$\sim$12-13\,\AA~(Fig.~\ref{fig:stellar}), and the
theoretically possible (achievable only through a 10\,Myr old
instantaneous starburst) maximum is $\sim$18\,\AA, this puts a firm
theoretical upper limit of $<$30\% on any non-stellar contribution,
and realistically makes anything larger than a few percent
unlikely. And since we do not see a localised dip in $W_{CO\,2-0}$
around the AGN positions, we conclude  
that any hot dust emission associated with the AGN is
completely obscured at near-IR wavelengths.

Table~\ref{tab:lum} lists the observed and dereddened luminosities
measured within apertures of diameter 1\arcsec\, centered on the
northern and southern nuclei, from \cite{tecza00}, this work, and from
archival NICMOS data; and also those obtained by integrating
the luminosity profiles out to 250\,pc. We
included the NICMOS data to obtain a third independent measurement,
since our measurements of the observed luminosities and those of
\cite{tecza00} differ by a factor of two. We find the NICMOS data
agree with our measurements to within 10\%.
As can be seen, for \cite{tecza00}, the dereddening only alters the
measured luminosities by $\sim$10\%, whereas for us, the difference is
circa a factor of two.
This difference is most likely due to the fact that we applied a spatially
dependent, rather than single-valued, correction.

\begin{table*}
\caption{Integrated Luminosities of the Nuclei\label{tab:lum}}
\begin{tabular}{l c c c c}
\hline \hline
 & North & North & South & South \\
 & observed & dereddened & observed & dereddened \\ 
 & [$10^8L_\odot$] & [$10^8L_\odot$] & [$10^8L_\odot$] & [$10^8L_\odot$] \\
\hline
Apertures, \cite{tecza00} & 2.2 & 2.3 &  5.8 & 6.8 \\ 
Apertures, this work & 5.6 & 9.8 & 15 & 40 \\ 
Integrated profiles, this work  & ... & 11 & ... & 59 \\
Apertures, NICMOS & 5.3 & ... & 17 & ... \\ 
\hline
\end{tabular}
\tablebib{Apertures are all diameter 1\arcsec, the profiles are integrated out to a radius of 0.5\arcsec.}
\end{table*}

\section{Kinematic Centres and Black Hole Locations}
\label{sec:kin:BH}

In this section, we attempt to confirm the hypothesis that the black hole
locations do identify the centres of the progenitors,
by independently determining the locations of the kinematic centres from the stellar
velocity field. As we outline below, for a number of reasons locating the kinematic centres of the observed stellar rotation reliably and accurately is extremely difficult to do. 
Nevertheless, within the uncertainties, we find that the BH positions of \cite{max07} are consistent with the kinematic centres.

\cite{max07} determined the positions of the two AGN in NGC\,6240 by combining images
taken in the near-infrared, radio, and X-ray regimes. 
Postulating that the southern sub-peak of the northern nucleus as seen
in the K-band (`N1' in the notation of \citealt{gerssen04}) is
coincident with the position of the northern black 
hole, they superpose radio data to infer that the southern AGN
is located to the north-west of the 2.2\,$\mu$m peak of the southern nucleus.
The offset is explained in terms of dust obscuration,
which is supported by their 3.6\,$\mu$m images showing the southern
continuum peak to be coincident with the posited black hole location.
Our data also support the existence of higher obscuration at this
location: Fig.~\ref{fig:extinction} shows that the region suffering the greatest
extinction overlaps with, and extends to the northwest of, the
southern nucleus. 
We identify the AGN positions of \cite{max07} to an accuracy of better than $\pm$1 pixel (0.05\arcsec) on our data using the 2.12\,$\mu$m continuum features as reference points (as
shown in Fig.~\ref{fig:BHpositions}), and the angular separation of the VLBA radio sources (1.511$\pm$0.003\arcsec, \citealt{max07} and references therein). The uncertainties on the BH positions as identified on the \cite{max07} data are smaller than the uncertainty introduced by this translation onto our data.

We use kinemetry \citep{kraj06} to 
parametrize the stellar velocity field which, as Fig.~\ref{fig:stellar}
shows, clearly exhibits ordered rotation around each of the two nuclei
separately.
In many cases, the centre of a galaxy and its axis ratio (or inclination)
and position angle (PA) can be extracted straightforwardly from the
isophotes.
This is not possible for NGC\,6240 because the K-band isophotes are strongly
asymmetric with respect to the black hole locations found by
\cite{max07}. Hence we must use our analysis
of the velocity field to also fix these parameters simultaneously.

Kinemetry decomposes a velocity map $K(a,\psi)$ into a series of
elliptical rings that can be expressed as the sum of a finite number of terms:
\begin{displaymath}
 K(a,\psi) = A_0(a) + \sum_{n=1}^{N} A_n(a) \sin(n\psi) + B_n(a)\cos(n\psi).
\end{displaymath}
By minimizing particular terms, one can in principle find the
centre, position angle, and axis ratio of the velocity field.
Under the assumption that the velocity field is due to an axisymmetric
thin disc, these correspond directly to the equivalent parameters for
the galaxy (although one should bear in mind that, for example, with non-circular
orbits the kinematic and isophotal major axes may not coincide; and
for a geometrically thick system the axis ratio may not represent the
inclination).

The correct choice of kinematic centre minimises $A_2$ and $B_2$ (which
are also weakly dependent on ellipticity and PA) and to a lesser
degree $A_1$, $A_3$, and $B_3$ (see \citealt{kraj06} for a more
intuitive description of these parameters). 
The correct PA yields minimal values of $A_1$, $A_3$, and $B_3$; 
correct ellipticity minimises $B_3$.
For a range of different ellipticities and PA, we place
the kinematic centre at each pixel within a grid of 11$\times$11 pixel
centered on the AGN location and calculate the corresponding sum of
$A_2$ and $B_2$.
This results in a set of 2D\,maps of $A_2 + B_2$ for each combination
of PA and ellipticity.
The position of the minimum of each map then yields the best-fitting
kinematic centre for that specific combination of PA and
ellipticity. 
If the best-fit to the kinematic centre is the same for
different values of ellipticity and PA (i.e. a unique best estimate for
the kinematic centre exists), we then proceed to determine $A_1$ and
$A_3$ at this kinematic centre for a range of different position
angles, taking as the best estimate of the PA that value which results
in $A_1+A_3$ being minimal. 
Finally, the ellipticity is found by
analogously minimising the $B_3$ coefficient.

\begin{figure}
\begin{center}
\includegraphics[width=0.48\textwidth]{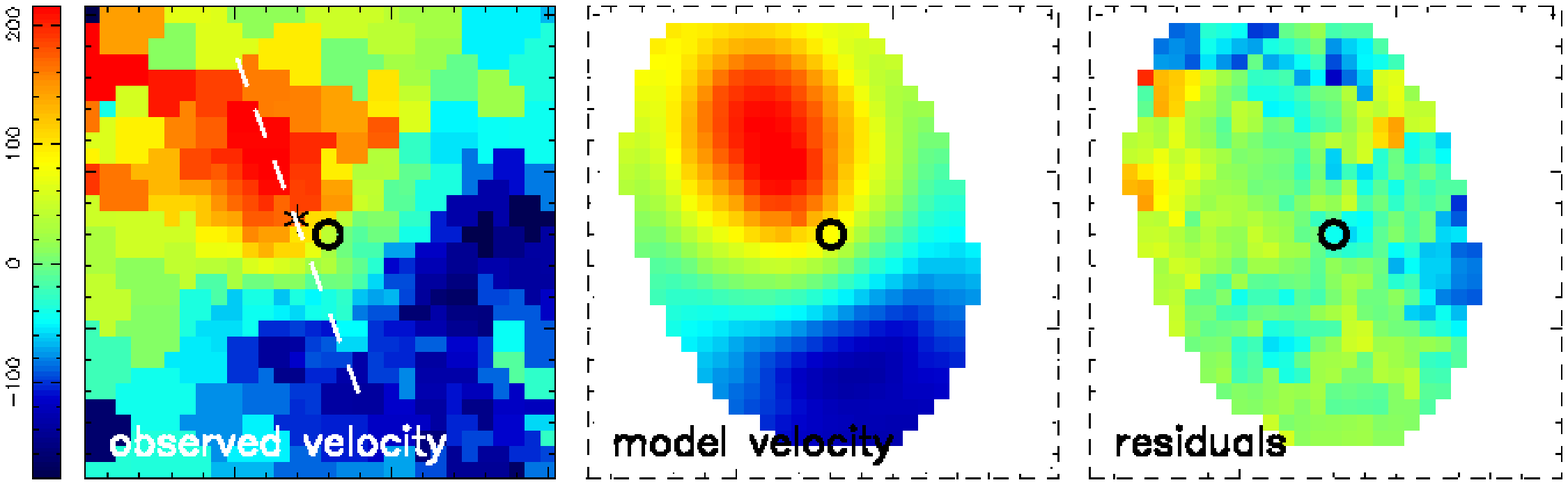}\\ 
\includegraphics[width=0.48\textwidth]{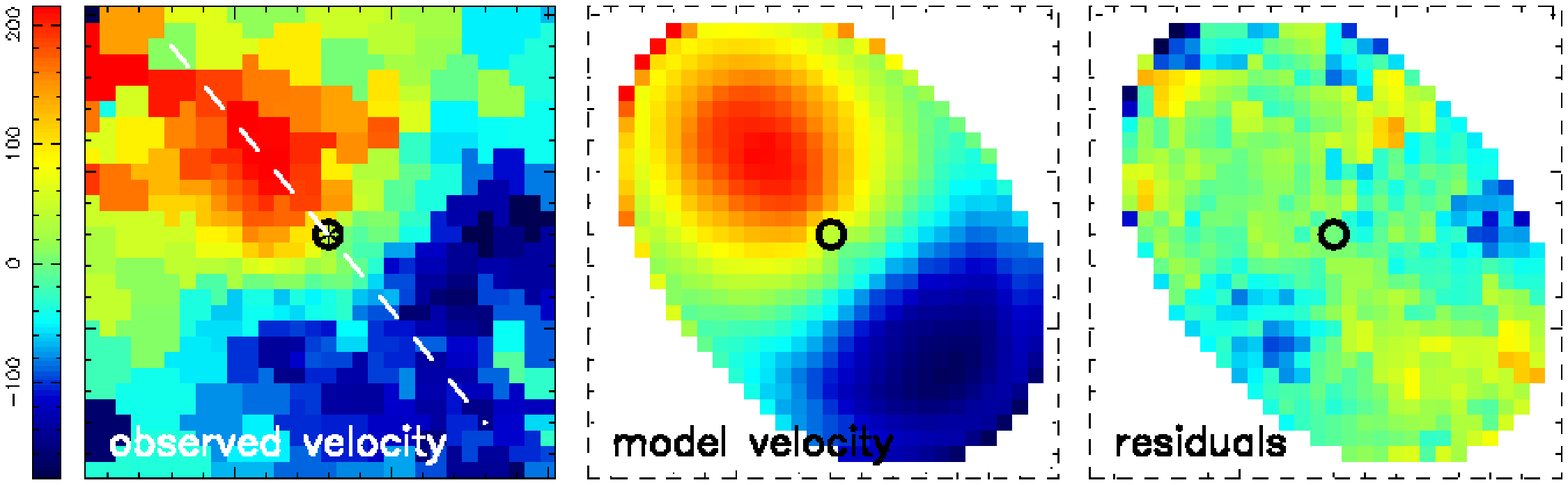}\\
\caption{Northern Nucleus: Comparison of kinemetry models for 2 different PAs and kinematic centres. It should be noted that the rotation curves do not differ significantly for the two different positions of the kinematic
  centre. Black circles indicate black hole positions; black stars represent the kinematic centre; the dashed white line traces the major axis of rotation. Left to right: velocity map, model, and residuals. First row: best-fitting kinematic centre from kinemetry analysis. The average residuals per pixel are 33.5\,km\,s$^{-1}$. Second row: kinematic centre fixed at the position of the AGN. The average residuals per pixel are 35.3\,km\,s$^{-1}$. Image size is 820$\times$720\,pc.}
\label{fig:starvelmodels:NN}
\end{center}
\end{figure}

\begin{figure}
\begin{center}
\includegraphics[width=0.48\textwidth]{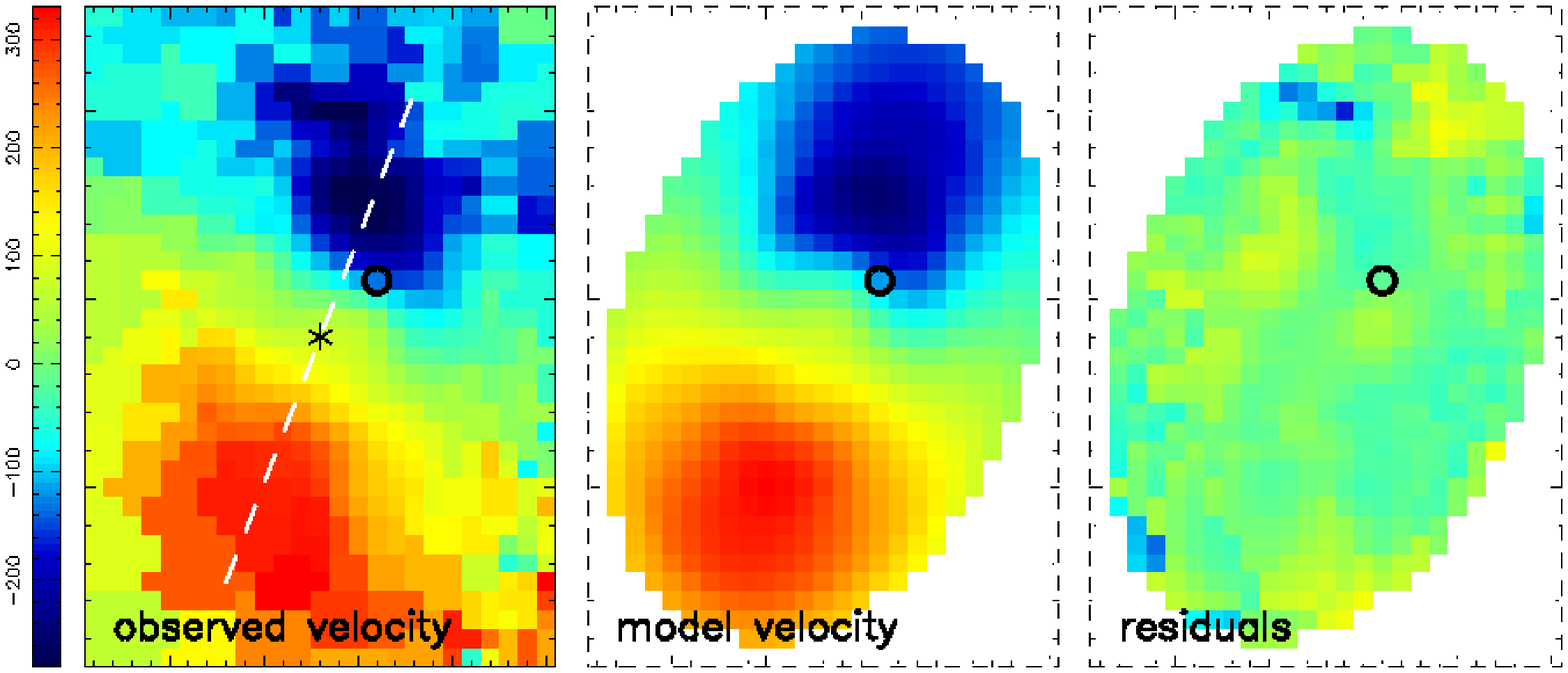}\\
\includegraphics[width=0.48\textwidth]{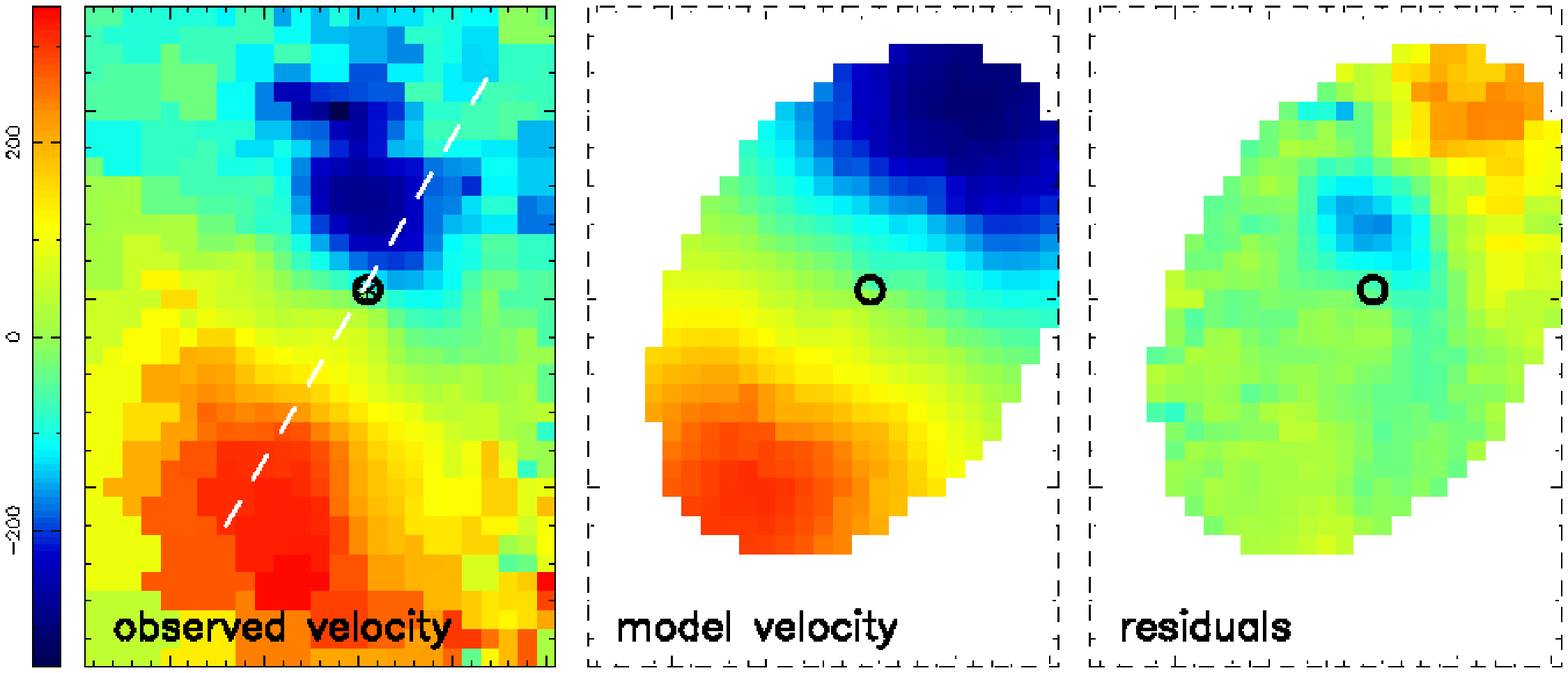}\\
\includegraphics[width=0.48\textwidth]{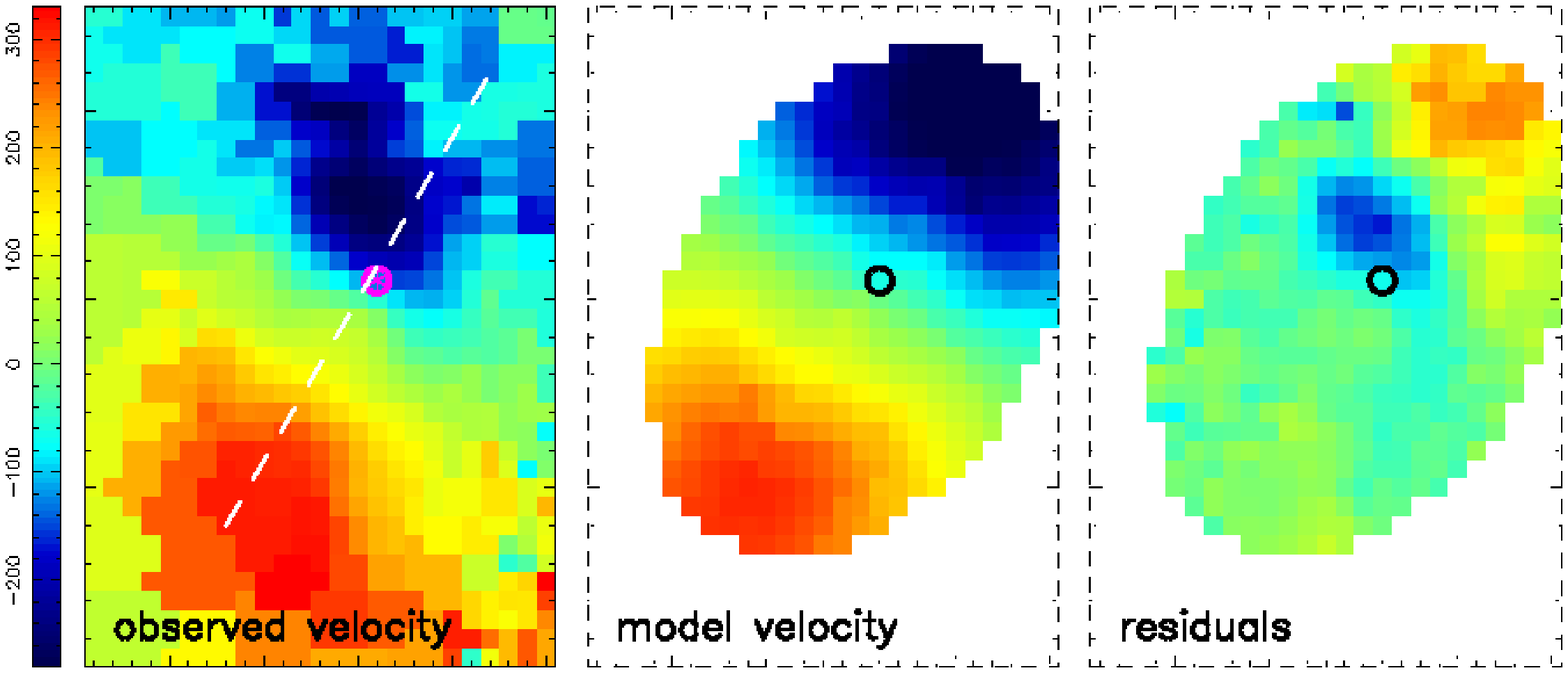}\\
\caption{Southern Nucleus: Comparison of kinemetry models for different PAs and kinematic centres. It should be noted that the rotation curves do not differ significantly for the two different positions of the kinematic
  centre. Black circles indicate black hole positions; black stars represent the kinematic centre; the dashed white line traces the major axis of rotation. Left to right: velocity map, model, and residuals. First row: best-fitting kinematic centre from kinemetry analysis. Average residuals per pixel are 28.2\,km\,s$^{-1}$. Second row: kinematic centre fixed at the position of the AGN; average residuals per pixel are 45.9\,km\,s$^{-1}$. Third row: kinematic centre fixed at position of the AGN and performing fit only to southern half of nucleus. Average residuals per pixel over southern half are 21.1\,km\,s$^{-1}$. Image size is 600$\times$840\,pc.}
\label{fig:starvelmodels:SN}
\end{center}
\end{figure}

With this method we are able to find a unique solution for all three
parameters for the southern nucleus.
But it fails for the northern nucleus where
the best estimates for kinematic centre and position angle are
interdependent: 
the minima for $A_2+B_2$ (determining the kinematic centre) 
lie on an arc with the exact position of the kinematic centre
dependent on the choice of position angle. 
The only way out of this impasse is to make additional assumptions, 
and so we arrive at a `best-fit' by assuming the velocity field to
have the same PA as the major axis of the continuum emission. 
We note that the PA of the continuum emission is independent of the
extinction correction, because the extinction and continuum maps have
the same PA.
Figures~\ref{fig:starvelmodels:NN} and~\ref{fig:starvelmodels:SN} show
the observed velocity field, the model, and the residuals for 2 cases
in each nucleus: when the best-fitting centre derived from the
kinemetry is used, and when the centre is fixed at the location of the
black hole.

In the northern nucleus, the location of the derived kinematic
centre is reasonably consistent with the black hole location,
differing by only 0.12\arcsec.
Since the average residuals per pixel are 33.5\,km\,s$^{-1}$ and
35.3\,km\,s$^{-1}$ for these two cases respectively, they can be
considered statistically indistinguishable.
We conclude that in the northern nucleus, we can confirm that the
black hole is located at the kinematic centre of the stellar rotation.

For the southern nucleus, the difference is more significant.
The separation between the best-fitting centre and the AGN location is
0.22\arcsec.
The mean residuals are 28.2\,km\,s$^{-1}$ (best-fit) and
45.9\,km\,s$^{-1}$ (AGN position). 
This can be understood with reference to the dispersion map in
Fig.~\ref{fig:stellar} and the extinction map in
Fig.~\ref{fig:extinction}.
The region to the northwest of the AGN in the southern nucleus
exhibits both anomalously high dispersion and extinction -- features
that are to be expected in a merger system.
It is therefore not clear whether the velocity field in this region is
really tracing the rotation of the progenitor, or something more
complex.
Inspection of the residuals in Fig.~\ref{fig:starvelmodels:SN} suggests
the latter, an issue that we discuss in more detail in
\S\ref{sec:stellkin}.
Our conclusion for the southern nucleus is that the centre we derive
from the velocity field is biased by perturbed kinematics on the
northern side of the nucleus.

In both nuclei, for further analyses we therefore adopt the black hole
locations of \cite{max07} as identifying the kinematic centres of the nuclei,
and make use of the rotation curves, dispersion, and luminosity
profiles centered at these positions.
We derive axis ratios of 0.70\,$\pm$\,0.02 and 0.65\,$\pm$\,0.02 and PAs of $229^\circ\pm2^\circ$ and $329^\circ\pm2^\circ$ (measured east of north, PA pointing from receding to approaching velocities) 
for the northern and southern nuclei, respectively.
We interpret the axis ratios in terms of an inclination for a flat
system, noting that this may be an overestimate since the nuclei are
likely to be thick. The impact is that we may also overestimate the
intrinsic velocity.

\section{Stellar Kinematics between the Nuclei}
\label{sec:stellkin}

NGC\,6240 is known to exhibit an exceptionally large stellar velocity
dispersion between the nuclei \citep{lester94,doyon94,tecza00}.
In Fig.~\ref{fig:stellar} we present a detailed 2D dispersion map
which shows that the region with the highest ($>$300\,km\,s$^{-1}$)
dispersion is fairly localised, and lies across the northern side of
the southern nucleus. 
The localised nature of this region, together with poor
signal-to-noise, most likely
explain why \cite{genzel01} and \cite{dasyra06b} cite a significantly
lower maximal dispersion, as it can easily be missed by slit
measurements.

This high dispersion cannot be associated with the established stellar
population in the southern nucleus because the high velocity
($\sim$500\,km\,s$^{-1}$ at 500\,pc) implies a dynamical timescale of 
$t_{dyn}$\,$\sim$5\,Myr.
Any asymmetry in the dispersion of a stellar population orbiting the
southern nucleus would be dispersed within this timescale.
An alternative explanation is that this represents a region where
emission from stars that have been formed recently as a result of the
interaction is superimposed on the light from the nucleus.
As the stellar kinematics are derived from absorption lines,
a superposition of two populations at different line-of-sight
velocities would lead to an overestimate of the dispersion: 
the line resulting from such a superposition would appear broader than the
intrinsic line widths of each of its constituent lines. 
Fitting the resulting line profile with a single kinematic component
would lead to a significant over-estimate of the disperion.
Support for this explanation is lent by \cite{ohy03}.
Based on a study of the morphology and kinematics of the warm and cold
gas, they argued that clouds along the line of sight to the
northern half of the southern nucleus were being crushed in the
interaction. 
In this scenario, if the cloud crushing leads to star formation, there
would indeed be two superimposed populations at this location.
This is supported by Spitzer observations of the Antennae galaxies, which show that the largest energy output, and thus a high rate of star formation, occurs in the overlap region between the two galaxies' nuclei \citep{brandl09}.

Although rather speculative, in the following discussion we consider
whether this scenario can in principle account for both the increased
dispersion and irregularities in the velocity in the northern half of
the southern nucleus.
Our aim is to use a very simple toy model to test the basic validity of
the hypothesis by reproducing the characteristic features of the data; we do
not attempt to match them exactly, nor to constrain the numerous
parameters that would be required to do so.

\begin{figure}
\begin{center}
\includegraphics[width=0.48\textwidth]{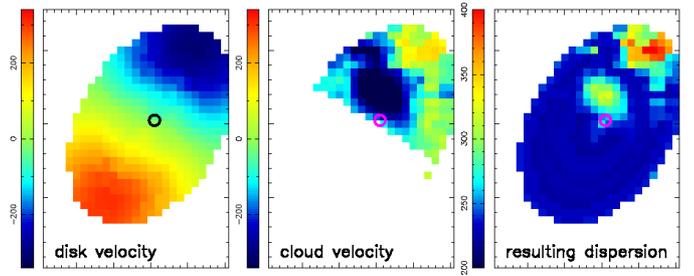}\\
\caption{Southern Nucleus: Resulting apparent dispersion of the
  superposition of a `disc' population and a `cloud-crushing'
  population, both with velocity dispersions of
  220\,km\,s$^{-1}$. Left to right: disc velocity field, cloud
  population velocity field, resulting apparent dispersion. Black
  circles indicate BH positions. The right panel shows that a simple
  `disc plus cloud' model can produce an increase in dispersion,
  characteristic of that observed (Fig.~\ref{fig:stellar}).
This indicates that the high dispersion in the northern half of the
  southern nucleus may be an artifact due to the presence of
  two kinematically different stellar populations with similar
  luminosities.}
\label{fig:starvelmodels:SN_model}
\end{center}
\end{figure}

We construct a two-population model as follows: 
the disc population is represented by a disc with the kinematic centre 
at the position of the AGN. PA, inclination, and velocity curve are found via a minimisation such as best to match the observed velocity field in the southern half. The dispersion is adopted from the radial dispersion curve derived from the southern half. 
We then subtract the disc velocity field from the observed velocity field in the northern half, and take this to be the velocity field of the second stellar population (thus ensuring that the observed velocity field is reproduced as closely as possible). In order to derive the resulting dispersion, we combine the two populations at each pixel by adding two Gaussians, with their centres corresponding to velocity and FWHM corresponding
to dispersion (resembling the superposition of two absorption or
emission lines). The resulting velocity field of the combination of
both populations is then extracted by fitting a Gaussian to the result
(resembling the method through which stellar kinematics are derived).
As can be seen in Fig.~\ref{fig:starvelmodels:SN_model}, the model, whilst not matching the observations exactly, does reproduce the characteristic features of the data, displaying locally significantly increased velocity dispersions. The fact that the dispersion map is not reproduced exactly is due to the fact that we intentionally kept the model simple, assuming spatially constant, equal weightings and line FWHMs.
We emphasise that the purpose of this exercise is simply to show that
it is in principle possible to account for the locally significantly
increased inferred dispersion through the effect of fitting a single
velocity and dispersion to an absorption feature to which two stellar
populations at different relative velocities have contributed. 
Although our assumptions that the two populations have the same
luminosity and dispersion are simplistic, 
this underlines the point that a significantly
better fit could be achieved if these parameters were allowed to vary.
However, we feel that pursuing this is unjustified given
constraints available from the data.

The total extinction-corrected luminosity emitted by the internuclear
area with $\sigma\ge330$km\,s$^{-1}$ is
$8.5\times10^8$\,L$_\odot$. 
If this can be attributed in
roughly equal shares to the old stellar population and the newly formed
stars, it would imply a stellar mass of $\sim$\,10$^8M_\odot$
contained in the starburst population in this area.
We speculate that this population may have originated from star
formation associated with gas between the nuclei, perhaps in the tidal
bridge or clump discussed in Section~\ref{sec:sims}.

\section{Jeans Modelling, Nuclear Masses, and Mass to Light Ratios}
\label{sec:kin:jeans}

Knowledge of the rotational velocity field allows one to make simple Keplerian
mass estimates. 
However, here the comparatively large velocity dispersions (generally
$\sigma/v_{rot}>1$) necessitate a more sophisticated approach
so as not to underestimate the actual mass. 
Two main avenues can be taken: Schwarzschild orbit superposition and
Jeans modelling. 
The former proceeds by reproducing the velocity and dispersion maps
from a large suite of stellar orbits calculated for a given potential,
which is varied until a match to the observations is achieved. 
The latter employs the Jeans equations, derived by taking
moments of the collisionless Boltzmann equation. 
For our data, Jeans modelling is better suited to recover the mass
distribution, as the accuracy and resolution of our kinematic data is
insufficient for Schwarzschild modelling.

A good overview of Jeans models can be found e.g. in \cite{BT08}. 
An important aspect of our approach is that
we separate the gravitational potential $\Phi$ from the distribution
of the stars, i.e. we allow for non-stellar mass. 
We assume isotropic velocity dispersion, which can be
justified because the data do not allow us to say anything about the form of
anisotropy, and this choice represents the least informative, and hence least 
constraining, option \citep{dejonghe86}. Another issue is the shape of the stellar and total mass distribution. A discy system would be characterised by $\sigma/v_{rot} \ll 1$, and
an oblate spheroidal system, in which the dispersion provides
non-rotational pressure support thickening the disc, by a
$\sigma/v_{rot} \gtrsim 1$.
Fig.~\ref{fig:stellar} shows that for both nuclei in NGC\,6240 
$\sigma/v_{rot} \sim 1$, suggesting that the latter is the more
physically realistic choice.
However, this is non-trivial to implement (see for example
\citealt{vanmarel07}).
Instead we make the simplifying assumption that both the stellar
distribution and potential are spherically symmetric,
which will allow us to capture the characteristics of the system.
It also has the advantage of yielding an analytical solution.

In order to assess the uncertainties of our result due to the choice of mass distribution, we also ran a set of models assuming an axisymmetric thin disc with an isotropic pressure component; the enclosed masses found with this model, which can be considered to be at the other end of the range of physically plausible models, agreed with the results of the spherically symmetric models within a factor of less than two. \cite{har06} model the gas kinematics of Cen A with three different Jeans models: an axisymmetric thin disc with and without pressure terms, and a spherically symmetric model. They find that the inferred black hole masses span less than an order of magnitude. Comparing their results with those of others, these authors find that, whilst favouring the pressure-supported thin disc model as the most physically plausible, the spherical Jeans model agrees best with the results from Schwarzschild orbit superposition modelling by \cite{silge05}.
The uncertainty of Jeans modelling results due to the unknown velocity
dispersion anisotropy is more difficult to assess due to the
degeneracy between the integrated mass and the anisotropy parameter
$\beta$. Both \cite{wolf09} and \cite{mamon09} attempt to quantify
this (see Fig.~1 in \citealt{wolf09} and Fig.~1 in \citealt{mamon09})
in the case of mass measurements for the Carina dSph and DM haloes,
respectively, finding that the range spanned by results for the
integrated mass assuming extreme values of $\beta$ are large at small
physical radii, decreasing to less than a factor of two at radii
comparable to or larger than the half-light radius.
Since a spheroidal system like the one we modelled
is unlikely to have extremely anisotropic velocity dispersions, and
since also we are measuring mass at radii larger than the half-light
radius, we can assume the factor of two derived by these authors to be an upper limit
on the error due to our choice of $\beta$=0. We thus estimate that the
uncertainties introduced by the assumptions inherent in our modelling
are unlikely to be larger than a factor of two.

Jeans modelling requires luminosity, rotational velocity, and velocity dispersion profiles as input.
Since the region between the two nuclei is likely strongly perturbed, we measured azimuthally averaged profiles using only a 180$^\circ$ wedge in the outside halves (i.e. opposite
the merger centre) of the galaxy, after subtracting the continuum associated with the recent starburst (assuming an intrinsic $W_{Br\gamma}$ of 22\,\AA, as measured at the knot of Br$\gamma$ emission north-west of the northern nucleus, cf. \S\ref{sec:starburst}) and correcting for extinction. To these we fitted a S\'ersic function to the profiles, constraining it with
the NIRC2 profile at $>0.4$\arcsec, and with the extinction-corrected
SINFONI profile at $<1.0$\arcsec\ (i.e. some overlap between the two
profiles is included). As part of the fitting process, the model profile was convolved in 2D
with the SINFONI PSF, to account for beam smearing at small scales. We then analytically deprojected these LOS luminosity profiles.
The measured rotation curves and dispersion profiles need to be corrected for beam smearing and projection along the line of sight in order to recover the intrinsic kinematics. 
For this, we used the code described in \cite{cre08}.

With these inputs, we compute M($r$). The total mass
enclosed within the cut-off radius (250\,pc and 320\,pc for northern
and southern nuclei, respectively) is found to be
$2.5\times10^9M_\odot$ and $1.3\times10^{10}M_\odot$. This is
comparable to that seen in the central few hundred parsecs of nearby
AGN (Figure 7, \citealt{dav07}).
We also calculate a global K-band mass-to-light ratio by finding the
multiplication factor that best matches the luminosity enclosed at
radius $r$ to M($r$). Here, as in the rest of the paper, L$_K$ is
taken to be the total luminosity in the 1.9--2.5$\mu$m band in units
of bolometric solar luminosity where
1\,L$_\odot$\,=\,3.8$\times$10$^{26}$W
(we note that a commonly used
alternative is the definition via solar $K$-band lumnosity density,
2.15$\times$10$^{25}$W\,$\mu$m$^{-1}$). 
We obtain values of
5.0\,$M_\odot/L_\odot$ and 1.9\,$M_\odot/L_\odot$ for the northern and
southern nuclei.
\cite{downessolomon98} find typical gas fractions of $\sim$15\% for local ULIRGs; in conjunction with
these modelling results, this implies total stellar masses of
$2.1\times10^9M_\odot$ and $1.1\times10^{10}M_\odot$, and stellar
mass-to-light ratios of 4.3\,$M_\odot/L_\odot$ and
1.6\,$M_\odot/L_\odot$ for the northern and southern nuclei,
respectively.
We discuss the possible inferences from these results regarding the nature of the nuclei and the progenitors' Hubble types in \S\ref{sec:SFhistory}.

It is tempting to compare the mass enclosed in the innermost few
parsecs to the expected black hole masses.
One avenue to estimate the black hole masses is afforded
by the X-ray luminosities. \cite{vig99} measure a total
absorption-corrected nuclear X-ray luminosity in the 2-10\,keV range
of 3.6$\times10^{44}$\,erg\,s$^{-1}$ with BeppoSAX. \cite{kom03},
using \textit{CHANDRA}, find that the northern and southern nuclei
have absorption-corrected 0.1-10\,keV X-ray luminosities of
0.7$\times10^{42}$\,erg\,s$^{-1}$ and
1.9$\times10^{42}$\,erg\,s$^{-1}$, respectively. 
The discrepancy is
most likely due to different absorption corrections (\citealt{kom03}
derive $N_H\sim10^{22}$cm$^{-2}$, \citealt{vig99} measure
$N_H\approx2\times10^{24}$cm$^{-2}$).
Since \cite{vig99} derive their
absorption correction from a significantly larger wavelength
range (up to 100\,keV) we use their values for the following
estimates. We convert the X-ray luminosity to monochromatic 5100\,\AA\
luminosity using the luminosity dependent $\alpha_{OX}$-relation
\citep{steffen06,maiolino07}, and then convert to AGN
bolometric luminosity adopting $L_{bol}$=7$\nu L_\nu$ (5100\AA)
\citep[e.g.][]{netzertr07}. 
\cite{dasyra06b} calculate Eddington
ratios for a sample of 34 ULIRGs, finding a median value of 0.37. With
this value for $L_{bol}/L_{Edd}$, we arrive at a combined black hole
mass of 4$\times$10$^8M_\odot$.
However, the uncertainty on this estimate is at least a factor of a few.


Alternatively estimates can be derived from the stellar
velocity dispersions. Using the M$_{BH}$-$\sigma$ relation from
\cite{tremaine2002} and the stellar dispersions measured at the
locations of the AGN ($\sim$200\,km\,s$^{-1}$ and
$\sim$220\,km\,s$^{-1}$ for the northern and southern nuclei,
respectively) yields expected central black hole masses of
1.4$\pm0.4\times$10$^8$M$_\odot$ and 2.0$\pm0.4\times$10$^8$M$_\odot$ for the
northern and southern nuclei, respectively.
The uncertainties on black hole masses derived from the
M$_{BH}$-$\sigma$ relation are also a factor of a few.

Finally, a measurement (with a rather smaller uncertainty)  of
$M_{BH}$ in the southern nucleus has recently been made using high
resolution stellar kinematics. 
Resolving the gravitational sphere of influence of the black hole and modelling the
kinematics, \cite{med10} derive a mass of $2.0\pm0.6\times10^9$\,M$_\odot$.

Although the range of $M_{BH}$ above spans an order of magnitude, the
very large uncertainties of the first 2 estimates means that all the
estimates are formally consistent (within $\sim$2$\sigma$).
We therefore caution against over-interpretation. 

\section{Merger Geometry \& Stage, And Cold Gas Concentration}
\label{sec:sims}

Starting from the assumption that the projected angular momenta of the nuclei are the true angular momenta, \cite{tecza00} proposed a merger geometry in which one of the nuclei is coplanar and prograde with the merger orbital plane, and the other is inclined with respect to the orbital plane (their Fig.~11). They further supported the notion of at least one of the galaxies being subject to a prograde encounter by noting that the formation of tidal tails such as those of NGC\,6240 is favoured in prograde encounters. Here we would like to build onto and expand this discussion.

NGC\,6240 indeed has long, well developed tidal tails, as can be seen
in optical and HI data \citep{gerssen04,hibbard01}, covering a
projected size of $\sim$\,45\,kpc from north to south, about half that
of the Antennae. In the HST images, a continuous dust distribution
across the centre strongly suggests that what we are seeing is an
extended, well developed single tail curving in front of the system,
rather than a series of disconnected shorter features. However,
NGC\,6240's tails are not as dense or long as those of e.g.~the
Antennae. What can we derive from this with regard to the merger
geometry? It is generally known that in merger simulations, the more
coplanar/prograde, and thus `resonant', an encounter is, the stronger
the resulting tidal tails are. It is quite clear however, from looking
at the relative orientation of the rotation axes of the nuclei, that
NGC\,6240 cannot be a perfectly coplanar merger. But a merger must not
be exactly coplanar and prograde in order to produce tidal tails -- an
example of this is NGC\,7252, which has tidal tails akin to those of
NGC\,6240. NGC\,7252's tidal features were successfully reproduced
through N-body simulations by \cite{hibbard95}; their simulation had
one of the progenitors on a coplanar, prograde encounter, but the
other galaxy at an inclination of $\sim$\,45\,degrees to the orbital
plane -- this illustrates that there is a non-negligible merger
parameter space around `coplanar/prograde' which will also result in
pronounced tidal tails. Therefore, whilst not pinning down the merger
geometry exactly, the existence of such long and well developed tidal
tails is convincing evidence that the merger geometry must be within a
reasonable parameter space around prograde/coplanar. The stellar
kinematics at first sight seem to disagree with this, giving the
appearance that the nuclei are counter-rotating. However, co-rotating
nuclei would also look like this if one nucleus is inclined behind the
plane of the sky, and the other in front. Thus the stellar kinematics
are consistent with the view that NGC\,6240 is not too far from being
a prograde merger. 
That the nuclei are clearly not exactly aligned could be the reason
for the additional shorter tails that make the system look rather
messy.
 
We therefore conclude that NGC\,6240's merger geometry most likely
tends towards coplanar/prograde, and in this sense is perhaps quite
similar to the geometry proposed for NGC\,7252 by \cite{hibbard95}.

It is more difficult to deduce the stage of merging the two galaxies
comprising NGC\,6240 are currently in - we cannot pin down their
merger geometry exactly, as discussed above, and neither do we know
the 3D relative velocities of the two nuclei, so we cannot calculate
whether the two nuclei will merge immediately or separate again before
eventually coalescing. We know that they must certainly have already
experienced their first close encounter, since tidal tails only form
after the first strong gravitational interaction of the two
galaxies. This is also supported by the high luminosity of the system,
because elevated levels of star formation also are only expected after
the first encounter leads to collision and compression of gas
complexes. And the fact that the two nuclei are still separated by
about a kpc in projection indicates that the system is not yet at the
`final coalescence' phase.

The observed peak of the gas emission between the nuclei
(Fig.~\ref{fig:co}), which \cite{tac99} found to display a velocity
gradient and interpreted as due to a self-gravitating gas disc located
in between the nuclei, is puzzling,
since from merger simulations this is generally not expected to
occur. We could be seeing a collapsed gas clump in a tidal arm - such
features are found to occur quite often in simulations of merging
galaxies above a certain gas fraction \citep{wetzstein07,bournaud08},
and are also seen in observations \citep{knierman03}.
An alternative, and perhaps more likely, interpretation of this gas
concentration is a tidal bridge connecting the two nuclei that is
viewed in projection. Such tidal bridges are often seen in merger
simulations. However, simulations have never yielded a system in which
the total gas mass is dominated by that between the nuclei.
And although bridges are produced most strongly in a perfectly
prograde coplanar encounter (which, as discussed above, is
inconsistent with our observations), as one deviates from a prograde
coplanar geometry the strength of the bridge decreases.

An alternative, although somewhat rarer, way to drive a significant gas mass away from the
nuclei is a direct interaction (i.e. low impact parameter), in which the ISM of the progenitors collides.
The best example of this is UGC\,12914/5 (the Taffy Galaxies;
\citealt{bra03}), in which about 25\% of the CO emission originates
between the nuclei.
In this system, the progenitors collided at about 600\,km\,s$^{-1}$.
\cite{bra03} argued that while this would have ionised the gas, the
cooling time is short enough that H$_2$ could reform while the discs
are still passing through each other.
From the ratio of $^{12}$CO to $^{13}$CO these authors found that the optical depth of CO in the bridge was far less than on the nuclei.
Combined with the effect of increased CO abundance due to grain destruction, this led them to suggest that the actual mass of molecular gas may be rather less than implied by the CO luminosity, perhaps only $\sim$6\% of the total gas mass.
Since direct collisions such as these tend to create ring galaxies (or in the case of the Taffy Galaxies, an incomplete ring or hook), this scenario seems unlikely for NGC\,6240.

In NGC\,6240, the CO map indicates that CO luminosity is dominated by that between the nuclei.
We must therefore consider physical effects that could
significantly affect the flux-to-mass conversion factor between the two
regions. It is already known that sub-thermal emission from
non-virialised clouds can radically modify the CO-to-H$_2$ conversion
factor, as shown in Fig.~10 of \cite{tacconi08}. Such an effect could
easily boost the emission from gas in a drawn-out, thin and diffuse
gas bridge compared to denser material around the nuclei. And it is
supported by other observations: a comparison of the CO line emission
(Fig.~\ref{fig:co}) with the 
1.315\,mm continuum measured by \cite{tac99} (their Fig.~3, left
panel) shows that the dust, as traced by the mm continuum, is
concentrated on the nuclei rather than following the CO line
emission. Furthermore, in a detailed analysis of various CO
transitions, \cite{gre09} were unable to find a single set of average
H$_2$ conditions that comes close to reproducing the observed line
ratios. A second important issue is the impact of X-ray irradiation on
the CO abundance around the nuclei, since the AGN are quite
luminous. \cite{lut03} argue that as little as 25--50\% of L$_{\rm
  bol}$ is due to the AGN. If we take the lower end of this range, and
distribute it equally between the two AGN, we find each radiates at
$L\sim3\times10^{44}$\,erg\,s$^{-1}$. We also adopt a column of
$2\times10^{24}$\,cm$^{-2}$ \citep{vig99} and assume a gas density of
$10^5$\,cm$^{-3}$. Under these conditions, hard X-rays from the AGN
will cause sufficient ionisation to reduce the CO abundance an order
of magnitude below its typical value of $10^{-4}$ out to a radius of
250\,pc (see \citealt{bog05}, and Davies \& Sternberg in prep.). Thus,
even if there is significant gas mass around the nuclei (as the models
imply), one would expect rather little CO emission. That these effects
will conspire to make the CO luminosity distribution look rather
different to the molecular gas mass distribution should be borne in
mind.

\section{Scale of Starburst \& Star Formation History}
\label{sec:starburst}

\begin{figure}
\begin{center}
\includegraphics[width=0.20\textwidth]{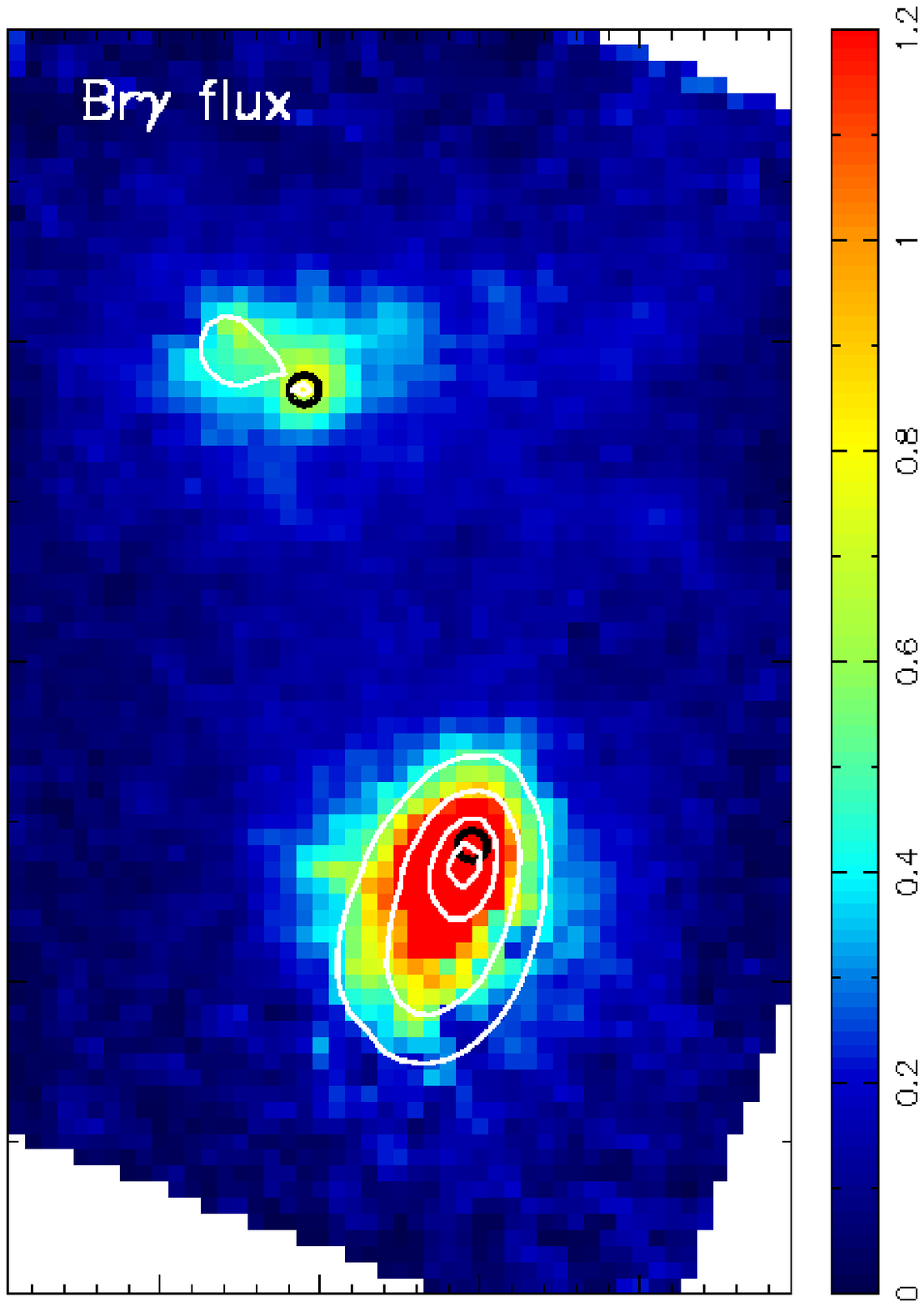}
\hspace{3mm}
\includegraphics[width=0.20\textwidth]{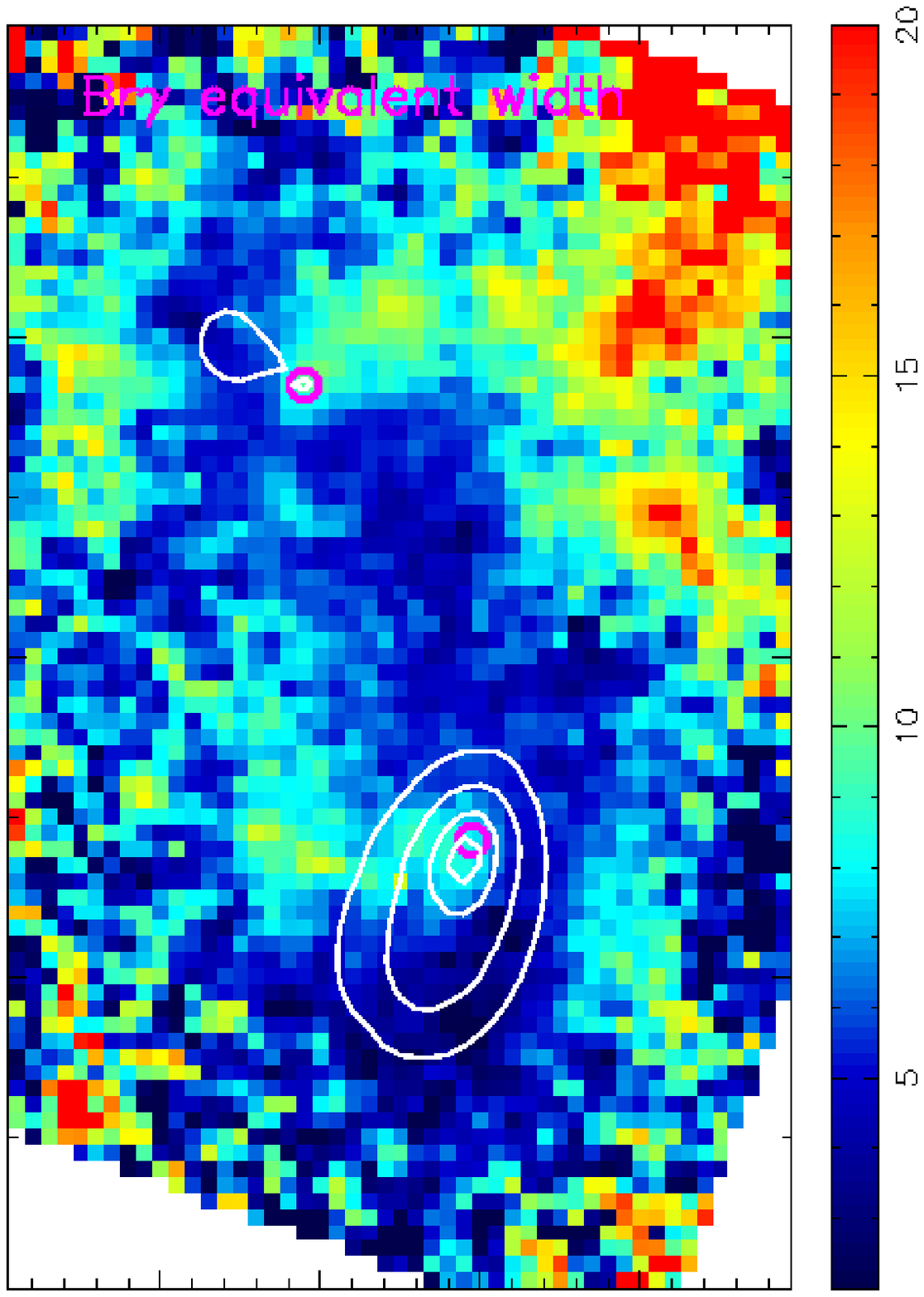}
\caption{Left: Map of Br$\gamma$ flux (units given in $10^{-16}$\,W\,m$^{-2}$\,$mu$m).
Right: Map of $W_{Br\gamma}$ (units given in \AA). White contours
tracing the K-band stellar continuum, and black circles denoting the
black hole locations are overdrawn on both maps. Although the most
intense Br$\gamma$ emission is on the nuclei, the equivalent width
here is lowest, suggesting that the K-band light is dominated by an
older stellar population. 
We note that in \S\ref{sec:extinction} we showed that dilution by hot
dust emission associated with the AGN cannot play a role in reducing $W_{Br\gamma}$.}
\label{fig:BrG}
\end{center}
\end{figure}

While \cite{tecza00} concluded that the nuclei were the progenitor
bulges, they did not quantify how much of the K-band
luminosity arises in the recent starburst and how much is due to the
progenitor bulges themselves.
Our data enable us to resolve this issue. 
Our method is to measure an average value for $W_{Br\gamma}$ away from
the nucleus where dilution from the bulge is small, and assert that
this is representative of the whole starburst.
The fundamental assumption is that, integrated over a sufficiently large
aperture, we are summing a fair cross-section of star
clusters and hence probing the average star formation properties.
This minimises the impact of the stochastic nature of individual
clusters.
We have therefore chosen a region containing a high density of clusters, and
measured $W_{Br\gamma}$ in a 0.8\arcsec\ aperture corresponding to
400\,pc so that many clusters are included.
Such a region lies about 1\arcsec\ to the west of the northern nucleus
\citep{pol07}, sufficiently far that the K-band continuum from the
nucleus is very faint. 
Fig.~\ref{fig:BrG} shows that, despite the Br$\gamma$ flux being
higher on the nuclei, both here (and also in other off-nuclear
regions) $W_{Br\gamma} = 22$\,$\pm$\,6\,\AA.
We use it to obtain the fractional contribution of the young stellar population
in 1\arcsec\ diameter apertures centred on the
nuclei, by dividing the measured $W_{Br\gamma}$ by
this intrinsic value.
We then find the K-band luminosity emitted by the starburst population
to be $3.6\times10^8L_\odot$ (northern nucleus) and
$1.3\times10^9L_\odot$ (southern nucleus), implying
L$_{K,young}$/L$_{K,total}$ to be 0.36 (northern nucleus) and
0.32 (southern nucleus) -- i.e. only about 1/3 of the total
K-band luminosity of the nuclei is due to recent star formation.

This implies that the K-band luminosity of the
nuclei is dominated by a population of stars older than
20\,Myr. 
An important question is whether this is consistent with claims
that one needs supergiant templates in order to match the CO bandhead
depth \citep{les88,sug97,tecza00}.
In the following, we show that basing conclusions about the dominant
stellar population on a fit to a single template is very uncertain.
The tables in \cite{ori93} and \cite{for00} show that the equivalent
widths of the CO bandheads vary considerably between individual stars;
and it is also well known that higher metallicity populations have
deeper bandheads because the K-band continuum is dominated by cooler
stars.
The templates found by \cite{tecza00} and \cite{sug97} to provide the
best fit to the K-band absorption features are of K4.5\,Ib and K2.5\,Ib
stars, respectively: i.e. specifically K type sub-luminous supergiants.
Both authors infer from this that the dominant stellar population is
late type supergiants.
\cite{tecza00} go on to conclude that these were produced as a
result of a burst of star formation that lasted for 5\,Myr and occurred
$\sim$20\,Myr ago.
We have examined the \cite{tecza00} starburst scenario in detail using
the population synthesis code STARS, 
and found that it would lead to $\sim$\,60\% of the K-band luminosity being due to M
supergiants (cooler than 4000\,K) and only $\sim$\,3\% due to K
supergiants (temperatures 4000--4600\,K).
Thus the scenario of a $\sim$\,20\,Myr old starburst leads to two apparent contradictions:
How can M supergiants dominate the luminosity and yet
have absorption features too deep to match the spectrum? 
And how can K supergiants provide a good match to the spectrum and
yet contribute only an insignificant fraction of the luminosity? 
The resolution is simply that the galaxy spectrum
consists of many different types of stars, and a single template can
at best be characteristic of the sum of all these.
The important conclusion here is that while the composite
spectrum is best matched by a supergiant template, other types of
stars still make up a significant fraction of the near-infrared
continuum.

In addition to the current scale of the starburst, another key point
of interest are the past and future star formation
rates. \cite{dimatteo07,dimatteo08} and \cite{cox06,cox08} offer a large sample
of medium-resolution simulations comprehensively covering a wide
range of gas fractions, merger geometries, and mass ratios, as well
as employing different codes. Their synthesised results regarding
the qualitative evolution of the star formation during a merger are
that nearly all encounters are marked by at least two peaks in the
star formation rate; one at the first encounter, and one upon final
coalescence. However, the relative strength of the two peaks
differs, depending on the merger geometry; but on average the second
peak is stronger \citep{dimatteo07,dimatteo08}.
This is particularly evident for mergers that tend towards
coplanar/prograde geometries, as appears to be the case for NGC6240.

The aforementioned simulations were carried out at medium resolution,
and thus one might ask whether their results would change when going to
higher resolution simulations resolving the multiphase ISM. However,
as \cite{bournaud08} and \cite{teyssier10} show, the qualitative evolution of the
star formation rate during a merger remains largely unchanged, the
only significant difference is that star formation proceeds much more
effectively, resulting in predicted star formation rates a factor of
up to ten times larger than those predicted by lower-resolution
simulations \citep{teyssier10}.

We therefore conclude that NGC\,6240, being between first encounter
and final coalescence, most likely experienced its first peak in
star formation rate in the recent past (triggered by the first
encounter); has currently elevated levels of star formation compared
to a quiescent galaxy; and will experience another, likely stronger, peak in star
formation rate in the near future
when the galaxies coalesce. This is supported by the observed
Br$\gamma$ emission, which indicates that star formation must
currently still be on-going, and by measurements of cluster ages
\citep{pol07} which are found to be typically very young -- in a
population of clusters with a range of ages, it is the brighter ones
that are more easily detected, and because clusters fade quickly,
these will also inevitably be younger. We furthermore note that since
NGC\,6240 is already just below the canonical ULIRG threshold of
L$_{IR}$\,$\gtrsim$\,10$^{12}L_\odot$, it is safe to predict that it
will breach this threshold once the final starburst is triggered, and
will become a \textit{bona~fide} ULIRG.

In order to be able to make more quantitative analyses, we use STARS
to calculate a range of observables from two star formation histories,
which both display the generic features discussed above; an initial
peak at first encounter followed by a gradual rise or plateau, and a
final intense burst. One of these simulations is selected from a set
of simulations intended to reproduce the properties of the Antennae
galaxies \citep{karl08,karl10}, and the other one from the library of
\cite{2009ApJ...690..802J}. 
Neither are coplanar/prograde, although they tend towards it, and
both produce
well-developed tidal tails during the interaction. The rationale of
using star formation histories from two different simulations is to
provide an estimate of the uncertainties introduced by the quantitative
differences introduced by the exact simulation details. Since we only
know that NGC\,6240 must be between first encounter and final
coalescence, we use the full range of simulated properties between
these two points, as well as our measurement and
associated uncertainties of $W_{Br\gamma}$, as constraints to derive
the uncertainties in the star formation properties.

In Table~\ref{tab:starburst}, we have used these star formation
histories to calculate the mass of young stars formed in both scenarios,
as well as the current star formation and supernova rates.
To do so, we have applied a scaling so that the K-band luminosity
matches that observed for each nucleus, under the assumption that the 
evolution of the central SFR mirrors that of the global SFR.
The table shows that quantitatively the results for the two different star formation histories do not differ greatly -- in fact, the variation within each scenario over the possible time range is larger than the variation between the averages of the two cases.

One conclusion is that the bolometric luminosity
of the starburst from the two nuclei together is only 20--30\% of the
system's total. This follows in the same direction as \cite{lut03} that the recent
starburst only contributes part of the total luminosity. We estimate a somewhat lower fraction than \cite{lut03}, who constrain starburst contribution to be 50 to 70\%, possibly due to the fact that we are looking at the central region whereas they investigated the global luminosities.

An important concern is that our results appear to be inconsistent with
those of \cite{bes01} who find, based on the 1.4\,GHz continuum,
significantly larger star formation and supernova rates of
83.1\,M$_\odot$ yr$^{-1}$ and 1.33\,yr$^{-1}$. We argue that this is due to the different star formation histories adopted. We show that this can have a major impact on interpretation of the
data, and that accounting for it makes the radio continuum data consistent with our results.

We first address a minor correction, specifically a
$\sim$10\% increase in radio flux due to the larger aperture used by
\cite{bes01}, which included a substantial flux contribution from an
off-nuclear source not included in our apertures (`N2' in their
nomenclature, see their Fig.~2). 
A larger correction may be needed by analogy to Arp\,220 for which
\cite{rovilos05} examined the 18-cm lightcurves of
the supernovae.
They found the type IIn RSNe model insufficient and arrived at a
supernova rate that was a factor of $\sim$3
smaller than the $\sim$\,2\,yr$^{-1}$ often quoted. A similar
correction may be applicable to the rates inferred from radio
measurements for NGC\,6240.
This may be related to the third issue, which concerns the star
formation history. 
The formulae used by \cite{bes01} to
convert 1.4\,GHz luminosity to star formation and supernova rates
(based on work by \citealt{condon90,condon92,cram98}) were derived
empirically based on measurements of normal disc galaxies for which a
constant star formation rate is characteristic.
This implicit constant star formation rate is very different to the
increasing star formation rate found in merger simulations.
We show below that this can make as much as a factor of three
difference in the star formation and supernova rates that are derived
from observables.

We use STARS to estimate the star formation rate from a given K-band
luminosity for three different star formation histories. First,
following \cite{bes01}, we adopt continuous star formation for a duration of 20\,Myr, corresponding to that estimated by \cite{tecza00}, \cite{pas03},
\cite{pol07} for the most recent star burst.
In order to reach L$_K$=1.7$\times10^9L_\odot$, corresponding to the
total measured for the starburst population,
we require a constant SFR of 25\,M$_\odot$ yr$^{-1}$.
On the other hand, for the recently increasing star formation rate 
typical of the merger scenarios, we find star formation rates of $\sim$\,10\,M$_\odot$ yr$^{-1}$.
Thus we find a factor $\sim$\,2.5 difference in the SFR
required to reach the same K-band luminosity, depending on the star formation history.
We can estimate the supernova rates in a similar way:
For 20\,Myr of continuous star formation at 25\,M$_\odot$ yr$^{-1}$,
we calculate a supernova rate of 0.3\,yr$^{-1}$. 
In contrast, for our merger scenarios, we
find supernova rates of $\sim$\,0.13\,yr$^{-1}$.
Thus, much of the discrepancy between our supernova rate and the
(corrected) value from \cite{bes01} is due to the different star
formation histories adopted.

This illustrates an important \textit{caveat}:
the choice of star formation history can have a far-reaching impact 
on the interpretation of the observables. 
We note that the very similar derived stellar population parameters
for our two different merger star formation histories indicate that,
for mergers, using a `characteristic' star formation history (i.e. an
initial peak at first encounter followed by a gradual rise or plateau,
and a final intense burst) will yield much more reliable quantities than either instantaneous or continuous star formation models.

\begin{table*}
\caption{Characteristics of Starburst Populations\label{tab:starburst}}
\centering
\begin{tabular}{l l c c c c c}
\hline \hline
simulation & nucleus & $L_{K,young}$ & $L_{bol,young}$ & mass of young stars & SFR & SNR \\
 & & [$L_\odot$] & [$L_\odot$] & [$M_\odot$] & [$M_\odot yr^{-1}$] & [$yr^{-1}$] \\
\hline
`1' & North & 3.6$\times10^8$ & 2.9$^{+0.5}_{-0.7}\times10^{10}$ & 4.2$^{+1.3}_{-0.4}\times10^8$ & $2.3^{+1.8}_{-0.7}$ & 0.03$\pm0.1$ \\
`2' & North & 3.6$\times10^8$ & 3.1$^{+0.4}_{-0.3}\times10^{10}$ & 3.7$\pm0.3$\,$\times10^8$ & 2.2$^{+0.4}_{-0.3}$ & 0.03$\pm0.01$ \\
`1' & South & 1.3$\times10^9$ & 10.0$\pm2.0$\,$\times10^{10}$ & 1.5$^{+0.5}_{-0.1}\times10^9$ & 8.2$^{+6.8}_{-2.4}$ & 0.11$^{+0.02}_{-0.04}$ \\
`2' & South & 1.3$\times10^9$ & 11.0$^{+2.0}_{-1.0}\times10^{10}$ & 1.3$\pm0.1$\,$\times10^9$ & 7.8$^{+1.7}_{-1.1}$ & 0.10$^{+0.02}_{-0.01}$ \\
\hline
\end{tabular}
\tablebib{$L_{K,young}$ was measured in 1\arcsec\,diameter apertures using the 
dereddened continuum flux map and adopting an intrinsic
$W_{Br\gamma}$\,=\,22\,\AA\ (see \S\ref{sec:starburst}). Values  
for $L_{bol}$, stellar mass, and SFR were inferred from this together
with the star formation history from the simulations at all points
between first encounter and coalescence; given are the median value and the range sampled.}
\end{table*}

\begin{figure}
\begin{center}
\includegraphics[width=0.45\textwidth]{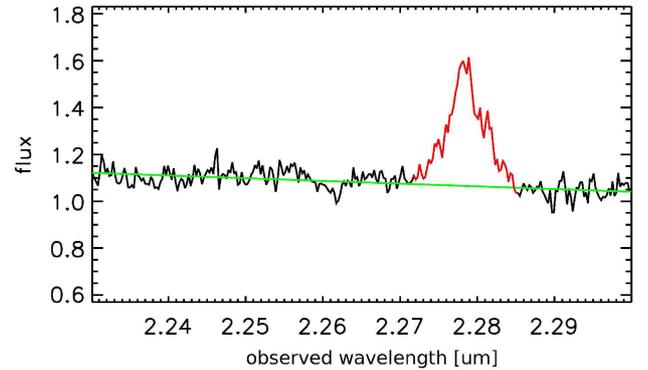}
\caption{Spectrum of the region to the northwest where the $W_{Br\gamma}$ is not diluted by continuum emission from the progenitors, integrated over an 0.8\arcsec\ diameter aperture. The flux is in units of 10$^{-16}$\,W\,m$^{-2}$\,$mu$m$^{-1}$. The measured $W_{Br\gamma}$ is 22\,$\pm$\,6\,\AA.}
\label{fig:EW_BrG}
\end{center}
\end{figure}

\section{Stellar Populations of NGC\,6240}
\label{sec:SFhistory}

\begin{table*}
\caption{Properties of the Nuclei\label{tab:bulgeprops}}
\centering
\begin{tabular}{c c c c c c c c c c}
\hline \hline
nucleus & v$_{max}$ & $\sigma$ & $n$ & $\epsilon$ & $r_{eff}$ & $r_{kin}$ & $L_K$ & $M_{dyn}$ & $M_{dyn}/L_K$ \\
 & [km\,s$^{-1}$] & [km\,s$^{-1}$] & & & [pc] & [pc] & [$L_\odot$] & [$L_\odot$] & [$L_\odot/M_\odot$] \\   
\hline
North & 205 & 200 & 2.8 & 0.3  & 490 & 265 & 1.1$\times10^9$ & 2.5$\times10^9$ & 5.0 \\ 
South & 300 & 220 & 2.8 & 0.35 & 180 & 290 & 5.9$\times10^9$ & 1.3$\times10^{10}$ & 1.9 \\ 
\hline
\end{tabular}
\tablebib{v$_{max}$ is the peak measured stellar velocity, $\sigma$ is the average/typical stellar dispersion (\S\ref{sec:obs:stellar}); $n$ and $r_{eff}$ are the S\'ersic index and effective radius of the light profile (\S\ref{sec:kin:jeans}), $\epsilon$ is the ellipticity (\S\ref{sec:kin:BH}), $r_{kin}$ is the extent to which rotation is seen, $L_K$ the total K-band luminosity (\S\ref{sec:extinction}), and $M_{dyn}$ the dynamical mass as derived from the Jeans modelling (\S\ref{sec:kin:jeans}).}
\end{table*}

\paragraph{Did the Progenitors Have Bulges?}
The existence or otherwise of bulges (where `bulge' here includes both pseudobulges and classical bulges) in the galaxies undergoing a merger has a profound impact on the star formation \citep{mih96}.
\cite{2005MNRAS.361..776S} show in their Fig.~15 that in a major merger of
discs (gas fraction 10\%) without
bulges, the first encounter already triggers an episode of enhanced
star formation that lasts for 300--400\,Myr.
The second encounter triggers a more intense, but also somewhat
shorter ($\sim$50\,Myr), starburst 800--900\,Myr later.
In a simulation that included a bulge and a central black hole, there was almost no perceptible change in the star formation rate at the first encounter.
This remained constant at $\sim$5--10\% of the peak star formation
rate until the second encounter, when the starburst is triggered and
the peak star formation rate was achieved.
Because there had been no earlier phase of enhanced star formation,
this starburst is more intense than it would be without a bulge.
Which of these two different scenarios holds true for NGC\,6240?

The evidence overwhelmingly points towards the progenitors having had a prominent bulge component.
First, the majority of massive spirals at low redshifts have bulges,
although at $z\gtrsim2$ the
situation may well be different as bulges may not yet have had time
to grow \citep{gen08}. Second, we know that both nuclei host an AGN
and therefore harbour a supermassive black hole.
The implication of the M$_{BH}$-$\sigma*$ relation is therefore that
the progenitors must have had bulges. 
Using estimates of the black hole masses based on the AGN luminosity and our measured stellar dispersion (\S\ref{sec:kin:jeans}) in conjunction with
the empirical correlation between M$_{BH}$ and bulge mass
\citep{haering04} yields associated bulge masses of
$8.7\times10^{10}M_\odot$ and $1.2\times10^{11}M_\odot$ for the
northern and southern nuclei respectively, an order of magnitude
larger than their dynamical masses. 
However, due to their transformational impact on both black hole and
host galaxy, it is uncertain whether such scaling relations can be
expected to hold during on-going mergers. 
An argument favouring the existence of bulges in the progenitors stems
from the merger simulation: the galaxies in our simulation had bulges,
and the good match to the observed $W_{Br\gamma}$ suggests that the
star formation history is reasonable.
As \cite{2005MNRAS.361..776S} and \cite{dimatteo07,dimatteo08} show, bulgeless mergers have markedly
different star formation histories, with the first encounter producing
the most potent burst of star formation (it should furthermore be noted that in higher-resolution simulations, the starburst triggered by the final coalescence occurs somewhat earlier; cf.~\citealt{teyssier10}.). Thus, at the current phase,
we would expect to see significantly lower $W_{Br\gamma}$ if the first
encounter had indeed triggered the more powerful star formation
event.

\paragraph{Are the Nuclei the Progenitors' Bulges?}
Are the nuclei we see (remnants of) the progenitor galaxies' bulges, did they originate only in the first encounter a few 100\,Myr ago, or is the rotation we see the signature of stellar discs? The latter is unlikely, since the high dispersion clearly indicates a thickened structure, and nuclear stellar discs in contrast are often accompanied by sigma-drops \citep{peletier07}. \cite{tecza00} proposed the first, based on an estimate of the global mass-to-light ratio of the nuclei. Our high spatial resolution data allow us to explore this in more detail using not only the mass and luminosity, but also the effective radius and kinematics, and our knowledge of the stellar populations.

Table~\ref{tab:bulgeprops} quantifies the kinematics, luminosities,
masses, and sizes of the two nuclei. How do these compare with the
characteristics expected of a bulge? 
An argument in favour of the nuclei being bulges is supplied by
estimates of the mass-to-light ratio of the non-starburst stellar
population: correcting the global stellar mass-to-light ratios
(\S\ref{sec:kin:jeans}, assuming a
gas fraction of $\sim$15\%) for the starburst contribution (Table~\ref{tab:starburst}), we find
that the non-starburst nuclear populations must have mass-to-light
ratios of $\sim$\,5.2\,$M_\odot/L_\odot$ (northern nucleus) and
1.8\,$M_\odot/L_\odot$ (southern nucleus) -- noting that the
associated uncertainties are estimated to be about a factor of two.
If the nuclei only formed after the first encounter ($\sim$560\,Myr ago in the `Antennae' scenario), they should have a M/L$_K$ of \textit{at most} 2\,M$_\odot$/L$_\odot$, whereas a several Gyr old stellar population has a M/L$_K$ of $\sim$6--10\,M$_\odot$/L$_\odot$ (e.g. Fig~4 in \citealt{dav07}). 
This, as well as the small relative contributions of the starburst population to the nuclear masses and luminosities, indicates that a substantial fraction of the nuclei consists of stars predating the beginning of the merger.

NGC\,6240 has often been cited as having anomalously high stellar
dispersion. However, excluding the region towards the northern half of
the southern nucleus which we discuss in \S\ref{sec:stellkin}, the
dispersions typical for the nuclei are $\sim$200\,km\,s$^{-1}$.
These are not unusual for late-type bulges, as Fig. 18 in
\cite{kormkenn04} shows. 
Placing the two nuclei on the $\epsilon$\,vs.\,$v_{max}/\sigma$
diagram of \cite{kormkenn04} (their Fig. 17), it appears that the
northern nucleus lies comfortably within the region typical for
late-type pseudobulges, whereas the southern nucleus would represent a
somewhat more extreme case, having a higher $v_{max}/\sigma$ than
other galaxies. In both cases, however, the kinematics strongly
support the case for a disc-like, secularly-grown pseudobulge rather
than a merger-formed classical bulge. The S\'ersic indices are rather
typical of classical bulges, but we caution that the luminosity
profiles are disturbed and influenced by recent star formation, and
hence the values derived from the S\'ersic fits should be treated with
caution. For the same reason, we prefer to measure the sizes from the
extents to which rotation can be seen (indicating the extent to which
the nuclei dominate the light distribution), rather than the effective
radii obtained from the S\'ersic fits; this gives radii of 265\,pc and
290\,pc for the northern and southern nuclei, respectively (as opposed
to 490\,pc and 180\,pc based on the S\'ersic fits). This places them
at the lower end but within the range given by \cite{mor99} who
measure the bulge radii for 40 spiral galaxies and find values ranging
from 230\,pc to 1.93\,kpc with an average of 650\,pc.
The same authors also look at the bulge masses, where they find values from 1.4$\times10^9$M$_\odot$ to 1.8$\times10^{11}$M$_\odot$ with an average of 2.5$\times10^{10}$M$_\odot$. Our analysis (\S\ref{sec:kin:jeans}) yields masses of 2.5$\times10^9$M$_\odot$ and 1.9$\times10^{10}$M$_\odot$, again at the lower end but within the range found by \cite{mor99}.

Would we expect the bulges to survive to this merger state? Comparatively little is known about the evolution of galaxy bulges during mergers. Observationally, \cite{aguero01} report velocity dispersions of 217\,km s$^{-1}$ and 280\,km s$^{-1}$ in the two nuclei of the merger system AM\,2049-691, and \cite{shier98} investigate 11 multiple-nucleus merging objects and find nucleus sizes between 200 and 1600\,pc (on average 600\,pc), velocity dispersions of 66-151\,km s$^{-1}$ (excluding NGC\,6240, which was also sampled) with an average of 109\,km s$^{-1}$. 
However, these were to a large part galaxies in earlier merger stages.
\cite{hernq93} simulates the merger of two spiral galaxies with bulges. His model suggests that the bulges remain very nearly intact until final coalescence. The spin (rotational velocity) is reported to remain nearly constant throughout the simulated merger.

In light of this and the fact that the nuclei are found towards the lower end of the range of masses and sizes found for a large sample of spiral galaxy bulges, we conjecture that the nuclei are indeed the progenitor galaxies' bulges, most likely pseudo-bulges, that have had their outer, less tightly bound layers stripped off during the merger, leading to the sub-average sizes. Dispersion may have increased due to `heating' during the interaction by gravitational impulse, and a gravitational potential deepened by influx of gas into the central region. Since the progenitors did not only have a prominent bulge component, but must also have had substantial discs in order to allow such pronounced tidal features as we observe to develop, the progenitors were most likely of Hubble type Sa/Sb.

\section{Conclusions}
\label{sec:sum}

We present new adaptive optics integral field spectroscopy near-IR data and CO(2-1) interferometric line observations of the nearby merger system NGC\,6240.
Our main conclusions are:
\begin{itemize}
\item
The \cite{calzetti00} reddening law provides the best fit to photometric data points spanning 0.45$\mu$m to 2.22$\mu$m. The spatially resolved extinction is generally moderate, with $A_K\sim2$--4.

\item
The locations of the stellar kinematic centres are consistent with the
black hole locations proposed by \cite{max07}.
However, an additional stellar population along the line of sight may
be perturbing the velocity field of the southern nucleus and causing
the higher dispersion across the northern side.

\item
Jeans modelling of the stellar kinematics, assuming spherical symmetry
and isotropic dispersion, gives $M=2.5\times10^9M_\odot$ 
and $M/L_K=5.0\,M_\odot/L_\odot$ for the northern nucleus
(out to 250\,pc), and 
$M=1.9\times10^{10}\,M_\odot$ and $M/L_K=5.0\,M_\odot/L_\odot$ for
the southern nucleus (out to 320\,pc).

\item 
The presence of tidal arms, and the still-separated nuclei indicate that NGC\,6240 must be between first encounter and final coalescence. The stellar velocity field, and strength and prominence of the tidal arms indicate that this system is not perfectly, but reasonably close to prograde coplanar.

\item
Although simulations can produce significant projected gas mass between the nuclei, these are never as prominent as the observed concentration in CO luminosity. In order to explain this, additional physical effects such as spatial differences in CO-to-H$_2$ conversion factor or CO
abundance are needed.

\item
Star formation histories from numerical simulations display generic features. Using these in combination with our constraints on the merger stage and measurements of $W_{Br\gamma}$, we calculate the properties of the young starburst population. We find that recent star formation accounts for only about 1/3 of the total K-band luminosity. Thus the stellar luminosity is
dominated by stars predating the merger.

\item
Whereas the differences between the starburst properties inferred from different merger star formation histories appropriate to NGC\,6240 are markedly small, adopting a constant star formation rate yields very different results (such as a actor 3 change in the derived SFR). We conclude that, when characterising the star formation properties of a merger, a `generic' merger star formation history should be adopted in preference to either instantaneous or constant star formation histories.

\item
After accounting for the recent star formation, the mass-to-light
ratios of the remaining stellar population are
5.2\,$M_\odot/L_\odot$ and 1.8\,$M_\odot/L_\odot$ for the northern and
southern nuclei respectively. This implies that a population of stars
older than $\sim$1\,Gyr contributes the majority of the nuclear
stellar masses and luminosities.
Combined with the measured size and $V/\sigma$, as well as results
from the simulations, it implies that the two nuclei 
are the remnants of the bulges of the progenitor galaxies.

\end{itemize}

\begin{acknowledgements}
The authors would like to thank the anonymous referee for a thorough and thoughtful reading of the manuscript, which helped to clarify and improve the paper.
H.~Engel would like to thank Scott Tremaine and Payel Das for
interesting and helpful discussions on Jeans modelling.
The numerical simulations were performed on the local SGI-Altix 3700 Bx2, 
which was partly funded by the Cluster of Excellence: ``Origin and
Structure of the Universe''. Some of the data presented here were obtained at the W. M. Keck Observatory, which is operated as a scientific partnership among the California Institute of Technology, the University of California, and NASA. This work was supported in part by the NSF Science and Technology Center for Adaptive Optics, managed by the University of California at Santa Cruz (under cooperative agreement no. AST-9876783).
\end{acknowledgements}

\end{document}